\documentclass[a4paper,fleqn,usenatbib]{mnras}
\usepackage{enumerate}
\usepackage{color}
\usepackage{graphicx}
\usepackage{amsmath}
\usepackage{multicol}
\usepackage{amssymb}
\usepackage{savesym}
\usepackage{color}
\usepackage{tipa}
\usepackage{csvsimple}
\usepackage{arydshln}
\usepackage{arydshln}

\newcommand{\perbeam}{\,{\rm beam}$^{-1}$}

\newcommand{\micro}{\text{$\mu$}}
\newcommand{\milli}{\text{m}}

\newcommand{\GHz}{\text{GHz}}

\newcommand{\jansky}{\text{Jy}}

\newcommand{\ergs}{\text{erg s$^{-1}$}}

\newcommand{\E}[1]{\times10^{#1}}
\newcommand{\Msol}{{\rm M}_{\odot}}

\newcommand{\Pl}[2]{$#1\pm#2$}

\newcommand{\flux}{\text{erg cm$^{-2}$ s$^{-1}$}}

\title[Counterparts of two eclipsing ULXs]
  {Multiband counterparts of two eclipsing ultraluminous X-ray sources in M\,51}
  
\author[R. Urquhart et al.]{
R. Urquhart,$^{1}$\thanks{Emails: ryan.urquhart@icrar.org (RU), rsoria@bao.ac.cn (RS)}
R. Soria,$^{2,1,3}$
H.M. Johnston,$^{3}$
M.W. Pakull,$^{4}$
C. Motch,$^{4}$\newauthor
A. Schwope,$^{5}$
J.C.A.~Miller-Jones$^{1}$
and G.E.~Anderson$^{1}$
\\
$^{1}$International Centre for Radio Astronomy Research, Curtin University, GPO Box U1987, Perth, WA 6845, Australia\\
$^{2}$National Astronomical Observatories, Chinese Academy of Sciences, Beijing 100012, China\\
$^{3}$Sydney Institute for Astronomy, School of Physics A28, The University of Sydney, Sydney, NSW 2006, Australia\\
$^{4}$Observatoire astronomique, Universit\'e de Strasbourg, CNRS, UMR 7550, 11 rue de l'Universit\'e, 67000, Strasbourg, France\\
$^{5}$Leibniz-Institut f\"ur Astrophysik Potsdam, An der Sternwarte 16, 14482, Potsdam, Germany
}

\date{Accepted XXX. Received YYY; in original form ZZZ}

\pubyear{2017}

\begin{document}
\label{firstpage}
\pagerange{\pageref{firstpage}--\pageref{lastpage}}
\maketitle

\begin{abstract}
We present the discovery and interpretation of ionized nebulae around two ultraluminous X-ray sources in M\,51; both sources share the rare property of showing X-ray eclipses by their companion stars, and are therefore prime targets for follow-up studies. Using archival {\it Hubble Space Telescope} images, we found an elongated, 100-pc-long emission-line structure associated with one X-ray source (CXOM51 J132940.0$+$471237; ULX-1 for simplicity), and a more circular, ionized nebula at the location of the second source (CXOM51 J132939.5$+$471244; ULX-2 for simplicity). We observed both nebulae with the Large Binocular Telescope's Multi-Object Double Spectrograph. From our analysis of the optical spectra, we argue that the gas in the ULX-1 bubble is shock-ionized, consistent with the effect of a jet with a kinetic power of $\approx$2 $\times 10^{39}$ erg s$^{-1}$. Additional X-ray photo-ionization may also be present, to explain the strength of high-ionization lines such as He {\sc ii} $\lambda 4686$ and [Ne {\sc v}] $\lambda 3426$. On the other hand, the emission lines from the ULX-2 bubble are typical for photoionization by normal O stars suggesting that the nebula is actually an H {\sc ii} region not physically related to the ULX but is simply a chance alignment. From archival Very Large Array data, we also detect spatially extended, steep-spectrum radio emission at the location of the ULX-1 bubble (consistent with its jet origin), but no radio counterpart for ULX-2 (consistent with the lack of shock-ionized gas around that source).

\end{abstract}


\section{Introduction} \label{intro}

Ultraluminous X-ray sources (ULXs) are off-nuclear, compact accreting X-ray sources with luminosities greater than those of Galactic stellar-mass black holes (see \citealt{2017ARA&A..55..303K} and \citealt{2011NewAR..55..166F} for reviews). They exceed the Eddington limit for a standard 10$\,\Msol$ black hole ($L_{\rm x} \gtrsim 10^{39}\,\ergs$) and are thought to be the result of either super-critical accretion onto stellar-mass compact objects or sub-critical accretion onto intermediate-mass black holes (IMBHs). Some of the brightest ULXs may contain IMBHs \citep{2009Natur.460...73F}; however, the consensus has moved towards super-Eddington stellar-mass accretors for most ULXs. The discovery of three neutron star ULXs \citep{2014Natur.514..202B,2016arXiv160907375I, 2016ApJ...831L..14F,2017ApJ...834...77F,2017MNRAS.466L..48I} has strengthened this interpretation. 

Being highly energetic sources, ULXs have a significant impact on their surroundings. Their extreme X-ray luminosity  photo-ionizes the surrounding gas. Examples of ULX nebulae with a dominant X-ray photo-ionized component are Holmberg\,II X-1 \citep{2002astro.ph..2488P} and NGC\,5408 X-1 \citep{2006MNRAS.368.1527S, 2003Sci...299..365K}. The ionizing X-ray luminosities inferred from the optical line emission of these nebulae match (within a factor of a few) the apparent X-ray luminosities of the compact objects, suggesting that at least those ULXs are not strongly beamed and their luminosities are intrinsically high ({\it e.g.} \citealt{2002astro.ph..2488P, 2003RMxAC..15..197P, 2009ApJ...697..950K}). In addition to high photon luminosity, there is theoretical and observational evidence of powerful outflows (winds or jets) in ULXs \citep{2005ApJ...628..368O,2011ApJ...736....2O, 2016Natur.533...64P}, consistent with their nature as super-critical accretors. The kinetic power of the outflows produces shock-ionized nebulae (ULX bubbles), such as those around IC\,342 X-1 \citep{2012ApJ...749...17C} or Holmberg IX X-1 \citep{1995ApJ...446L..75M, 2006IAUS..230..302G, 2006IAUS..230..293P}. Tell-tale signs of the presence of shock-ionized gas are the expansion velocity of the bubble ($\sim$100--300 km s$^{-1}$) and the high [S\,{\footnotesize II}]/H$\alpha$ line ratio; in fact, some ULX bubbles were previously mistakenly identified as supernova or hypernova remnants. In most cases, photo-ionized and shock-ionized gas co-exist in a ULX bubble; the origin of the ionizing photons may be both the direct emission from the central object and the forward shock precursor. 

In a small number of shock-ionized ULX bubbles there is also direct evidence of collimated jets instead of (or in addition to) wide-angle outflows. The evidence is the elongated morphology of the bubble with symmetrical hot spots or lobes/ears. In particular, this is seen in NGC\,7793-S26 \citep{2010Natur.466..209P,2010MNRAS.409..541S} and M\,83-MQ1 \citep{2014Sci...343.1330S}.
In both cases, the X-ray luminosity from the central source is less than the ULX threshold ($\approx$10$^{39}$ erg s$^{-1}$), either because our direct view of the source is mostly obscured, or because the instantaneous accretion rate is currently low. It is their kinetic power ($P_{\rm jet} > 10^{39}$ erg s$^{-1}$), inferred from the expansion velocity of the shocked gas and from the optical/IR line emission of the bubble \citep{2010Natur.466..209P}, that puts those three sources in the class of super-critical accreting stellar-mass objects. Another well-known ULX bubble, MF16 in NGC\,6946 \citep{2003MNRAS.341L..49R,1994ApJ...425L..77V}, has an elongated morphology consistent with the propagation of a symmetrical jet into the interstellar medium (ISM); in that case, the direct X-ray luminosity of the central source does exceed the ULX threshold, but may still be only a small part of the bolometric luminosity, mostly reprocessed in the UV band \citep{2010ApJ...714L.167K}. 

Evidence of powerful jets has also been found in a few other ULXs that do not have optically bright, shock-ionized hot spots/lobes. In the ultraluminous supersoft source M\,81 ULS-1, the jet is revealed by the red and blue Doppler shifts of its Balmer emission lines, corresponding to $v \approx$ 0.17c \citep{2015Natur.528..108L}. The M\,81 source is classified as an ultraluminous supersoft source, because of its thermal spectrum with blackbody temperature of $\approx$80\,eV \citep{2002ApJ...574..382S}. The presence of a collimated jet in that source suggests that jets can co-exist with thick winds, which are likely to be the hallmark of the supersoft class \citep{2016MNRAS.456.1837S,2016MNRAS.456.1859U,2007MNRAS.377.1187P}; however, no evidence of jets has been found so far in other sources of that type. 

More often, it is bright radio emission with 5-GHz luminosity $\nu  L_{\nu} \ga 10^{34}$ erg s$^{-1}$ that indicates the presence of a jet or at least of fast outflows. The reason is that some of the bulk kinetic energy of the outflow is used to accelerate relativistic electrons at the reverse shock; compression of the magnetic field lines in the shocked ISM leads to electron cooling via synchrotron radiation \citep{1984RvMP...56..255B}. In Holmberg\,II X-1, the jet is revealed by a flaring radio core and radio-bright ``knots" of synchrotron emission either side of the core \citep{2014MNRAS.439L...1C, 2015MNRAS.452...24C}. The three extragalactic sources with elongated optical bubbles mentioned above (NGC\,7793-S26, M\,83-MQ1, NGC\,6946-MF16) all have strong optically-thin synchrotron emission. By analogy, the equally strong, optically-thin radio nebula (diameter $\approx$ 40 pc; \citealt{2007ApJ...666...79L}) around NGC\,5408 X-1 is also interpreted as evidence of a jet. The strong radio source associated with NGC\,5457 X-9 \citep{2013MNRAS.436.1546M} could be another example, although in that case it is difficult to disentangle the ULX contribution from the surrounding H{\footnotesize {II}} region.
In the Milky Way, SS\,433 very likely belongs to the same class of jetted, super-critical sources with extended bubbles and mechanical power $\ga$10$^{39}$ erg s$^{-1}$ \citep{2017MNRAS.467.4777F,2011MNRAS.414.2838G,2007A&A...463..611B,2004ASPRv..12....1F,1980MNRAS.192..731Z}; in that source, tell-tale signatures of the large-scale jet impact onto the ISM are the radio ``ears" protruding out of the W\,50 bubble \citep{1998AJ....116.1842D} and the characteristic optical line emission from filaments of shock-ionized gas \citep{2007MNRAS.381..308B}.

Not every optically-bright, shock-ionized ULX bubble has a radio counterpart (NGC\,1313 X-2 is a classical example of a large, optically bright, radio-quiet ULX bubble), and not every ULX with a bright radio counterpart is surrounded by an optically-bright, shock-ionized bubble (as we said for Holmberg\,II X-1). For a given kinetic power in the jet, the composition of the jet (leptonic or baryonic), the collimation angle, the nature of the compact object, the gas density and magnetic field strength in the ISM, and the age of the bubble may affect the fraction of kinetic power that goes into synchrotron cooling. In other cases, such as the radio flares detected in the transient M\,31 ULX \citep{2013Natur.493..187M}, the jet may be active only for a short fraction of time during state transitions; this is in contrast with sources such as NGC\,7793-S26 where the average jet power is $\sim$ several 10$^{40}$ erg s$^{-1}$ over $\sim$10$^5$ yrs \citep{2010Natur.466..209P}.

\begin{figure}
\centering
\includegraphics[width=0.46\textwidth]{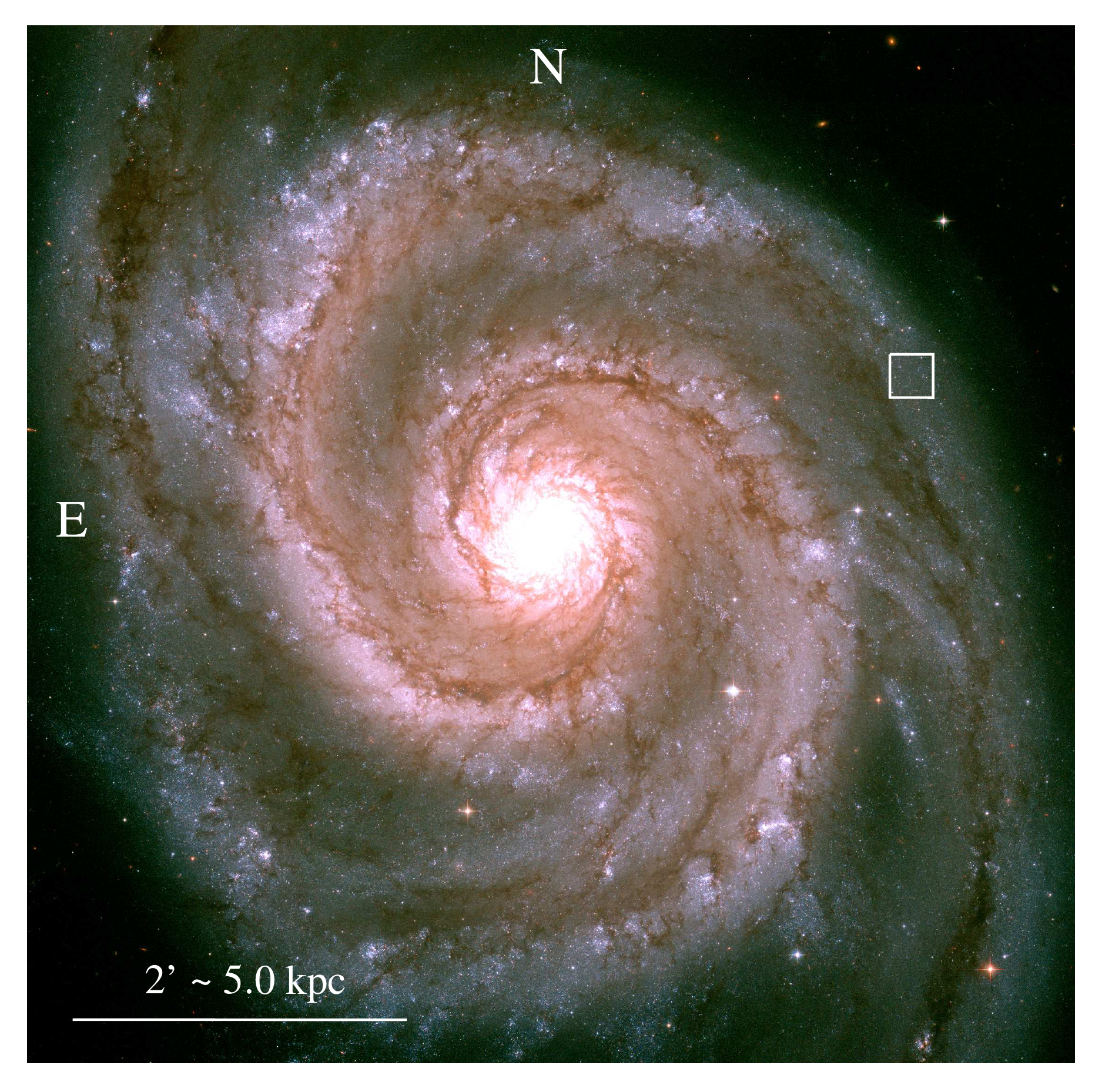}
\includegraphics[width=0.46\textwidth]{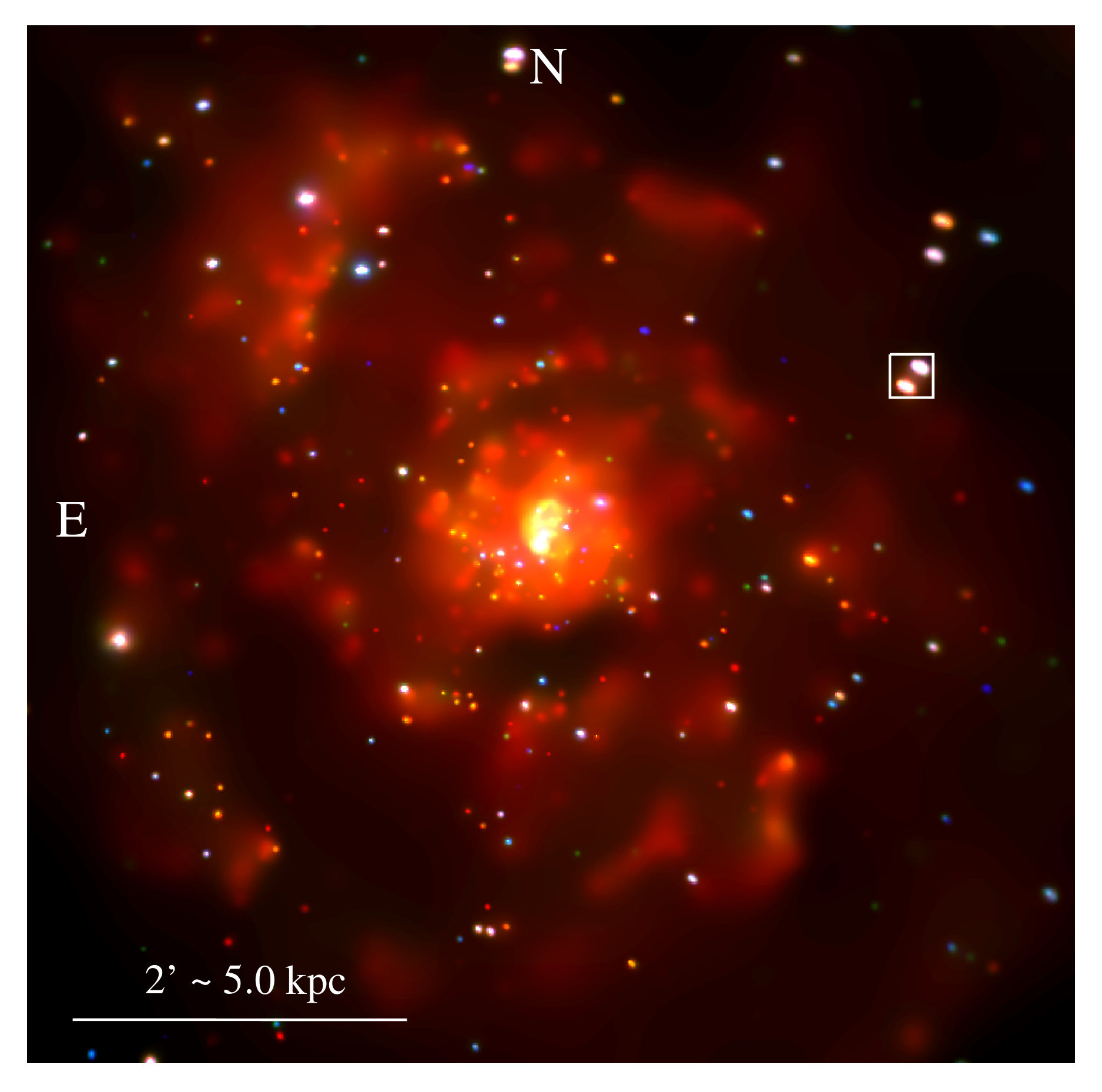}
 \caption{Top panel: {\it HST}/ACS-WFC RGB colour image of M\,51. Blue represents the F435W filter, green represents F555W and red represents F814W. The 15$^{\prime\prime}\times$15$^{\prime\prime}$ white box contains ULX-1 and ULX-2 and shows the region in which we performed PSF photometry. Bottom panel: {\it Chandra}/ACIS X-ray colour image, on the same scale as the {\it HST} image. Red $= 0.3$--$1$ keV, green $= 1$--2 keV, and blue $= 2$--7 keV. Inside the box, ULX-1 is the lower source (softer colour) while ULX-2 is the upper source (harder colour).}
  \label{M51_im}
  \vspace{0.3cm}
\end{figure}

We are conducting a study of ULXs associated with ionized nebulae in nearby galaxies, trying to determine which of them have evidence of jets, and whether the presence of a jet correlates with some X-ray spectral properties of the central source. We have recently identified \citep{2016ApJ...831...56U} two ULXs in the same spiral arm (Figure \ref{M51_im}) of the interacting galaxy M\,51 ($d=8.58\pm 0.10$\, Mpc; \citealt{M51dist_new}). The most intriguing property of those two sources is that both of them show sharp X-ray eclipses, likely due to occulations by their respective companion stars. For the first source, M\,51 ULX-1, we were able to determine that the period is either $\approx$6--6.5 days, or $\approx$12--13 days; we did not detect enough eclipses to identify the period of the second source, M\,51 ULX-2, but we estimated $P\sim10$ days based on its eclipse fraction \citep{2016ApJ...831...56U}. Both ULXs have similar luminosities ($L_{\rm X} \approx$ a few $10^{39}\,\ergs$), must be viewed at high inclination (being eclipsing sources), and are likely to be located in ISM regions with similar properties (as they are only $\sim 400$ pc apart along the same arm); however, they have different X-ray spectral characteristics \citep{2016ApJ...831...56U}. In this paper, we show that both ULXs are surrounded by ionized gas. In one case (M\,51 ULX-1), the ULX is coincident with an elongated ``jet-like" bubble. The other ULX (M\,51 ULX-2) is surrounded by a quasi-circular bubble. We will argue that in ULX-2 the nebula is consistent with an ordinary H\,{\footnotesize{II}} region photo-ionized by young stars, while in ULX-1, the gas is shock-ionized and the morphology suggests a jet origin. We will also show that the candidate jet source has a radio counterpart, while the other ULX does not. 



\section{Observations and Data Reduction}

\subsection{Archival {\it HST} images}

Publicly available {\it Hubble Space Telesecope} ({\it HST}) data for M\,51 were downloaded from the Hubble Legacy Archive\footnote{http://hla.stsci.edu/hlaview.html}. The observations were part of the {\it HST} Mosaic of M\,51 (proposal ID 10452) taken on 2005 January 12 with the Advanced Camera for Surveys (ACS), Wide Field Channel (WFC). This dataset includes three broadband filters (F435W, F555W, F814W) and the narrowband filter F658N, which covers H$\alpha$ and the two adjacent [N\,{\footnotesize{II}}] lines (central wavelength of 6584\AA{} and effective width of 75\AA{}). The total exposure time for the F814W and F555W bands was 8160 s, while for F435W it was 16320 s and for F658N 15640 s. The archival data were processed ``on-the-fly" using the standard ACS pipeline (CALACS: \citealt{2000ASPC..216..433H}), correcting for flat-fielding, bias and dark current. We then built a normalized, weighted average of the images in the F814W and F555W bands, and subtracted it from the F658N image, in order to remove the continuum stellar contribution from the narrow-band filter. Henceforth, when we refer to the F658N image, we mean the continuum-subtracted image unless stated otherwise. 

\begin{figure}
\centering
\includegraphics[width=0.48\textwidth]{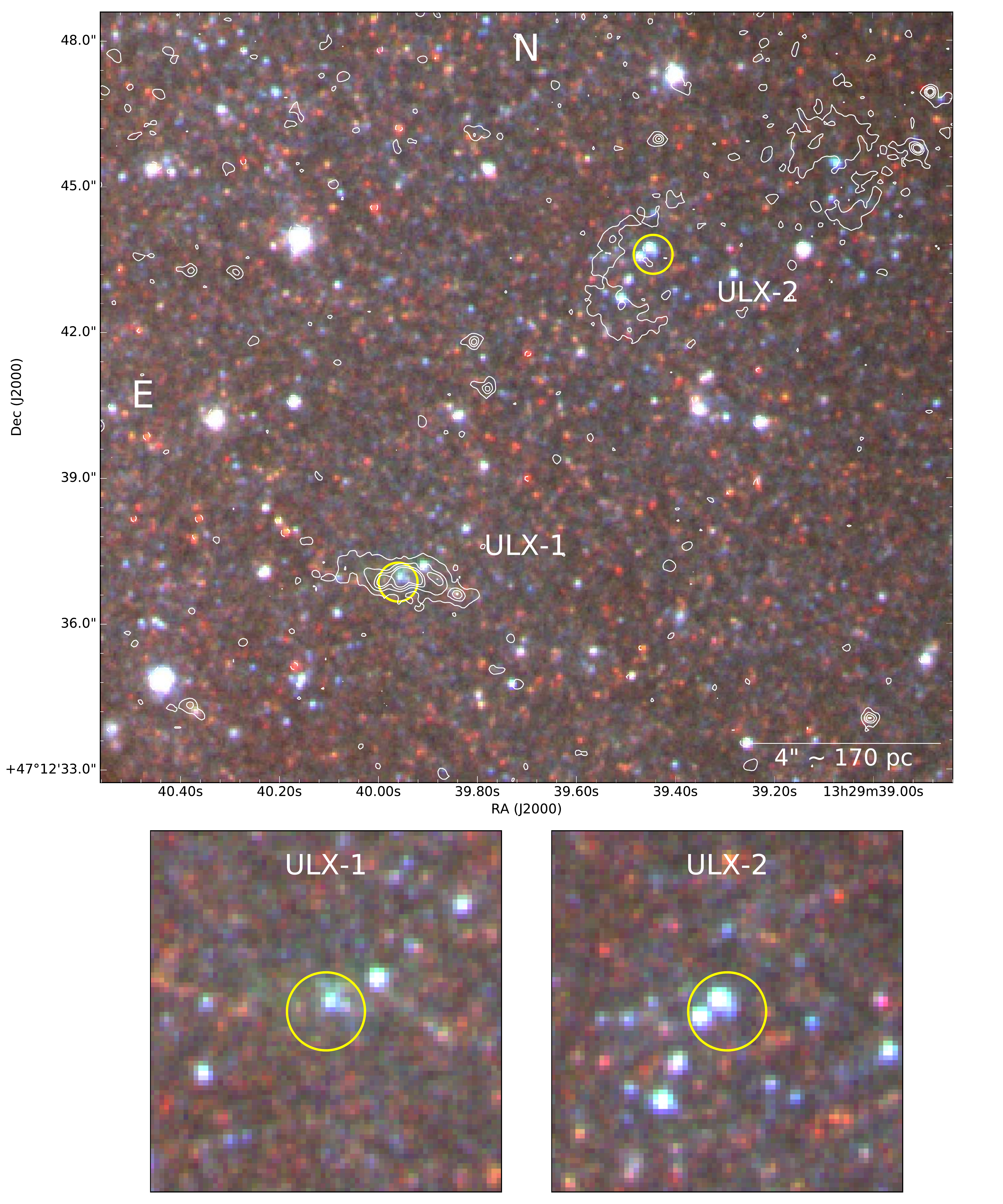}
 \caption{Top panel: zoomed-in {\it HST}/ACS image of the white box displayed in Figure \ref{M51_im}. Blue represents the F435W band, green $=$ F555W and red $=$ F814W. Yellow circles represent the {\it Chandra} positions of ULX-1 and ULX-2 with 0$^{\prime\prime}$.4 errors. White contours represent optical line emission in the continuum-subtracted F658N band. Line emission can be seen near both ULXs; in ULX-1 there is an elongated jet-like structure while ULX-2 is inside a more spherical nebula. Contours levels are arbitrary and are simply used to indicate the morphology of the ionized gas nebulae. Bottom left panel: zoomed-in view of the ULX-1 field. Bottom right panel: zoomed-in view of the ULX-2 field.}
  \label{field_im}
  \vspace{0.3cm}
\end{figure}

We performed aperture and point-spread function (PSF) photometry on the F435W, F555W and F814W filters, using the {\sc daophot} package \citep{1987PASP...99..191S} embedded within the Image Reduction and Analysis Facility ({\sc iraf}) software Version 2.16 \citep{1993ASPC...52..173T}. The {\it daofind} task was used to identify point-like sources within a 50$^{\prime\prime}\times$50$^{\prime\prime}$ region centred on the mid-point between ULX-1 and ULX-2 (a concentric region approximately four times larger than the white box in Figure \ref{M51_im}). Aperture photometry was then performed on these stars using {\it phot}. Due to the crowded nature of the field (Figure \ref{field_im}), a circular source aperture with a radius of 3 pixels (0$^{\prime\prime}$.15) was used; the encircled energy fraction within that radius, for a point-like source, is $\approx$80\%. For the PSF photometry, several bright, isolated sources were manually selected and inspected (with the tasks {\it pstselect} and {\it psf}); each candidate PSF star was manually inspected and those that appeared slightly extended were rejected. The average PSF created from those stars was then used as input to the photometry task {\it allstar}. PSF photometry was conducted on all stars within a 15$^{\prime\prime}\times$15$^{\prime\prime}$ box centred on the mid-point of ULX-1 and ULX-2 (white box overplotted in Figure \ref{M51_im} and displayed in its entirety in Figure \ref{field_im}). In addition to the photometry of point-like sources, we measured the total background-subtracted count rates of the emission nebulae around the two ULXs. Finally, count rates of stellar and nebular sources were converted to physical magnitudes and fluxes in the various bands, using the zeropoint tables for ACS-WFC available on the STScI web site\footnote{http://www.stsci.edu/hst/acs/analysis/zeropoints/old\_page/\\localZeropoints}.

\subsection{New LBT spectra}

\begin{figure}
    \centering
    \includegraphics[width=0.48\textwidth]{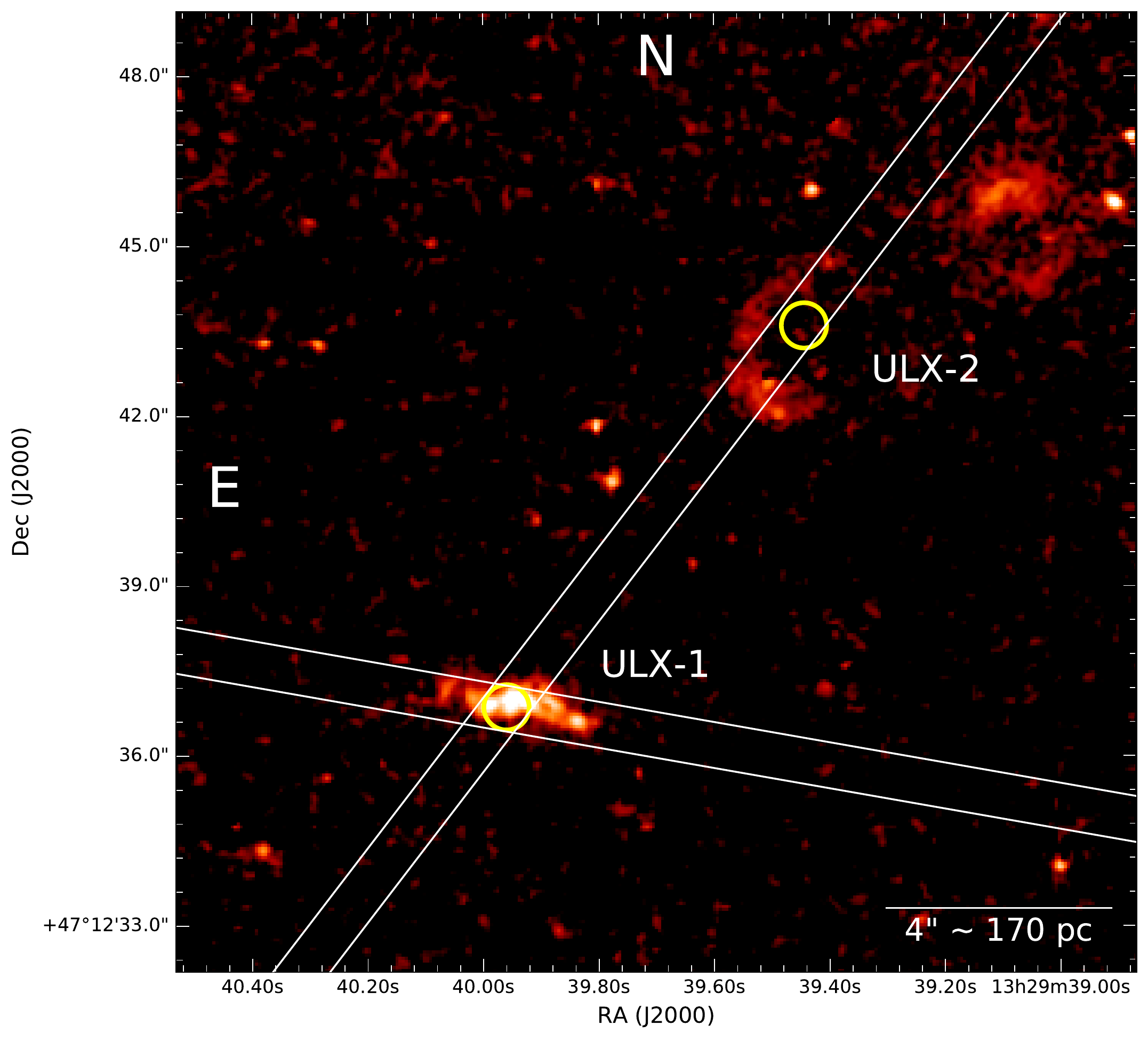}
    \caption{LBT slit positions, overplotted on the continuum-subtracted F658N image. In the first observation, the slit was aligned along the jet-like structure near ULX-1, including also the candidate optical counterpart of the ULX. This corresponded to an instrument celestial position angle of 80$^{\circ}$ East of North. The second slit was positioned so as to observe the candidate optical counterparts of both ULX-1 and ULX-2, along with some of the diffuse emission around the two X-ray sources. This corresponded to a position angle of $322.5^{\circ}$. The yellow circles represent the {\it Chandra} positions of ULX-1 and ULX-2 with 0$^{\prime\prime}$.4 uncertainty.}
    \label{LBT_slit_im}
\end{figure}

Optical spectroscopic data together with the corresponding calibration exposures (dark and bias frames, flat fields, arcs) were taken with the Large Binocular Telescope's (LBT) Multi-Object Double Spectrograph (MODS) \citep{2010SPIE.7735E..0AP}. The targets were observed from 2016-05-10 UTC 06:31:53.064 (MJD 57518.272142) for a total of 3$\times$900=2700 seconds, on each of the two MODS, and for each of two slit positions. We used a 0$^{\prime\prime}$.8-wide, segmented long slit with gratings G400L and G670L for the blue and red beam channels, respectively. The blue G400L grating has a resolution of 1850 at 4000\AA{} and a nominal dispersion of 0.5\AA{} per pixel, while the red G670L has a resolution of 2300 at 7600\AA{} and nominal dispersion of 0.8\AA{} per pixel. In the first observing configuration (OB1), the slit was oriented along the jet-like structure of ULX-1 (position angle PA $= 80^{\circ}$, measured from North to East); in the second configuration (OB2), the slit was oriented to contain both the (candidate) point-like optical counterparts of ULX-1 and ULX-2 (Figure \ref{LBT_slit_im}), at PA $= 322.5^{\circ}$.

The raw data were bias-subtracted and flat-fielded with the reduction package modsCCDRed\footnote{http://www.astronomy.ohio-state.edu/MODS/Software/\\modsCCDRed/} Version 2.0.1, provided by Ohio State University. Spectral trimming and wavelength calibration were done with the Munich Image Data Analysis System ({\sc midas}; \citealt{1992ASPC...25..115W}). We then used standard {\sc iraf} tasks for further analysis. We extracted one-dimensional background-subtracted spectra with the {\sc iraf} task {\it apall}. We flux-calibrated the red and blue spectra using the response files provided by the MODS instrumentation team. Finally, we analysed the spectra using the {\sc iraf} task {\it splot}; in particular, for each emission line, we measured equivalent width (EW), full width at half maximum (FWHM), and central wavelength (by fitting a Gaussian profile). 

One of the main parameters we were interested in was the total H$\alpha$ flux from the elongated nebula around ULX-1. The LBT slit was placed and oriented in order to cover as much of that region as possible (Figure \ref{LBT_slit_im}) but some of the flux (about 20 percent) falls out of the slit. However, we can measure the total emission of that nebula from the {\it HST} F658N image, which includes the flux from H$\alpha$ plus [N {\sc ii}] $\lambda$6548 and [N {\sc ii}] $\lambda$6583. We combined the information on the H$\alpha$/[N {\sc ii}] line ratio obtained from the LBT spectra, with the measurement of the total H$\alpha$ $+$ [N {\sc ii}] flux from the {\it HST} image. This gave us an absolute measurement of the total H$\alpha$ luminosity of the nebula, regardless of what fraction of it falls on the LBT slit. After calibrating the H$\alpha$ flux, we then obtained the fluxes of all other emission lines in both the red and the blue spectrum, using their ratios to H$\alpha$. (In other words, we assumed as a first-order approximation that the same fraction of total flux was collected in the red and blue spectra.) We applied a similar procedure to determine the total H$\alpha$ flux (and by extension, the fluxes in the other lines) of the nebula around ULX-2, which could only be partly covered by the 0$^{\prime\prime}$.8 LBT slit (Figure \ref{LBT_slit_im}). Finally, we de-reddened the spectra assuming line-of-sight extinction $A_{\rm V}=0.095$ mag corresponding to $E(B-V) = 0.031$ mag \citep{2011ApJ...737..103S}, using the extinction curves of \citet{1989ApJ...345..245C}. We chose to correct only for the line-of-sight Galactic extinction as a simple, model-independent, standard reference value. We are aware that there must additional extinction intrinsic to M\,51, but it is difficult to quantify it. The total dust reddening of the nebulae is likely less than or similar to the value indirectly inferred from X-ray spectral fitting of the two ULXs inside the nebulae \citep{2016ApJ...831...56U}, that is $E(B-V)_{\rm int} \la 0.1$ mag; a plausible guess is that the intrinsic extinction through the M\,51 halo is similar to the extinction through the Milky Way halo, that is the total $A_{\rm V}$ is about twice the Galactic line-of-sight value.

An additional caveat is that MODS does not have an Atmospheric Dispersion Corrector, so the spectral data suffer from slit losses due to atmospheric dispersion, when the slit is not fixed at or near the parallactic angle. Normalizing the H$\alpha$ line flux to the value inferred from the {\it HST} image allows us to sidestep this problem, at least for the red spectra. We used the Ohio State University's online calculator\footnote{http://www.astronomy.ohio-state.edu/MODS/ObsTools/\\obstools.html\#DAR} to estimate the effect that differential atmospheric refraction may have on our MODS spectra. The observations with the slit in the OB1 configuration were taken at an hour angle $\approx 0^h.9$--$1^h.2$; the differential refraction between H$\alpha$ and H$\beta$ is $\la$0$^{\prime\prime}$.2. For the OB2 configuration, the spectra were taken at hour angles $\approx 2^h.2$--$2^h.5$, corresponding to differential refractions $\approx$0$^{\prime\prime}$.3 between H$\alpha$ and H$\beta$ and $\approx$0$^{\prime\prime}$.4 between H$\alpha$ and H$\gamma$. This shift is already of the same size as the typical spatial scale of structural inhomogeneities (brighter and fainter clumps) in the two nebulae; therefore, in principle, a fraction of nebular emission may be on the slit at H$\alpha$ but fall outside the slit at bluer wavelengths. We will argue a posteriori, from our results, that the uncertainty on the total extinction and on the effect of differential refraction are small and do not affect our interpretation.


\subsection{Archival VLA data}

Radio maps of M\,51 were provided by G. Dumas (private communication); see \citet{2011AJ....141...41D} for a detailed discussion of the instrumental setup. In summary, observations were taken from 1998 to 2005 using the Very Large Array (VLA) at 1.4\,GHz (20\,cm) in the A, B, C and D configurations, at 4.9\,GHz (6\,cm) in the B, C and D configurations, and at 8.4\,GHz (3.6\,cm) in the C and D configurations. A total integration time of $\sim170$ hr over all three frequencies was achieved. The 8.4 and 4.9\,GHz observations had been combined with data from the Effelsberg 100-m telescope \citet{2011AJ....141...41D}, to detect extended emission regions that are resolved out in the VLA maps. The synthesised beam sizes for the 8.4\,GHz, 4.9\,GHz and 1.4\,GHz are 2$^{\prime\prime}$.4, 2$^{\prime\prime}$.0 and 1$^{\prime\prime}$.5 respectively.   
Radio fluxes were determined using the Astronomical Imaging Processing System ({\sc aips}; \citealt{2003ASSL..285..109G}). We fitted elliptical Gaussian profiles in the image plane, using the task \texttt{imfit} to determine the integrated flux of the detected sources; we also calculated peak fluxes and root-mean-square noise levels in the field around the ULXs. 
We placed 3-$\sigma$ upper limits on non-detections.

\subsection{Astrometric alignment}

In order to identify any optical and/or radio candidate counterparts of ULX-1 and ULX-2, we first needed to verify and, if necessary, improve the astrometric alignment between the {\it Chandra}, {\it HST}, and VLA images. For our first iteration, we took the default {\it Chandra} astrometry of the re-processed and stacked archival data. The X-ray data were processed and analyzed with standard tasks within the Chandra Interactive Analysis of Observations ({\sc ciao}) Version 4.7 \citep{2006SPIE.6270E..1VF}; for a description of the {\it Chandra} observations, see \cite{2016ApJ...831...56U}. The {\it Chandra}/ACIS-S astrometry is known to be accurate within $\approx$0$^{\prime\prime}$.5 for 68\% of the observations, and within $\approx$0$^{\prime\prime}$.7 for 90\% of the observations\footnote{http://cxc.harvard.edu/cal/ASPECT/celmon/}. The {\it HST} astrometry was aligned onto the 2MASS catalogue \citep{2006AJ....131.1163S}, which provides an accuracy within 0$^{\prime\prime}$.3. The VLA astrometry was assumed absolutely correct by default. With this first set of astrometric solutions, we looked for coincidences between the X-ray, optical, and radio sources (not including the ULXs that were the subject of our study), within $\approx$0$^{\prime\prime}$.5. 

\begin{figure}
\centering
\includegraphics[width=0.48\textwidth]{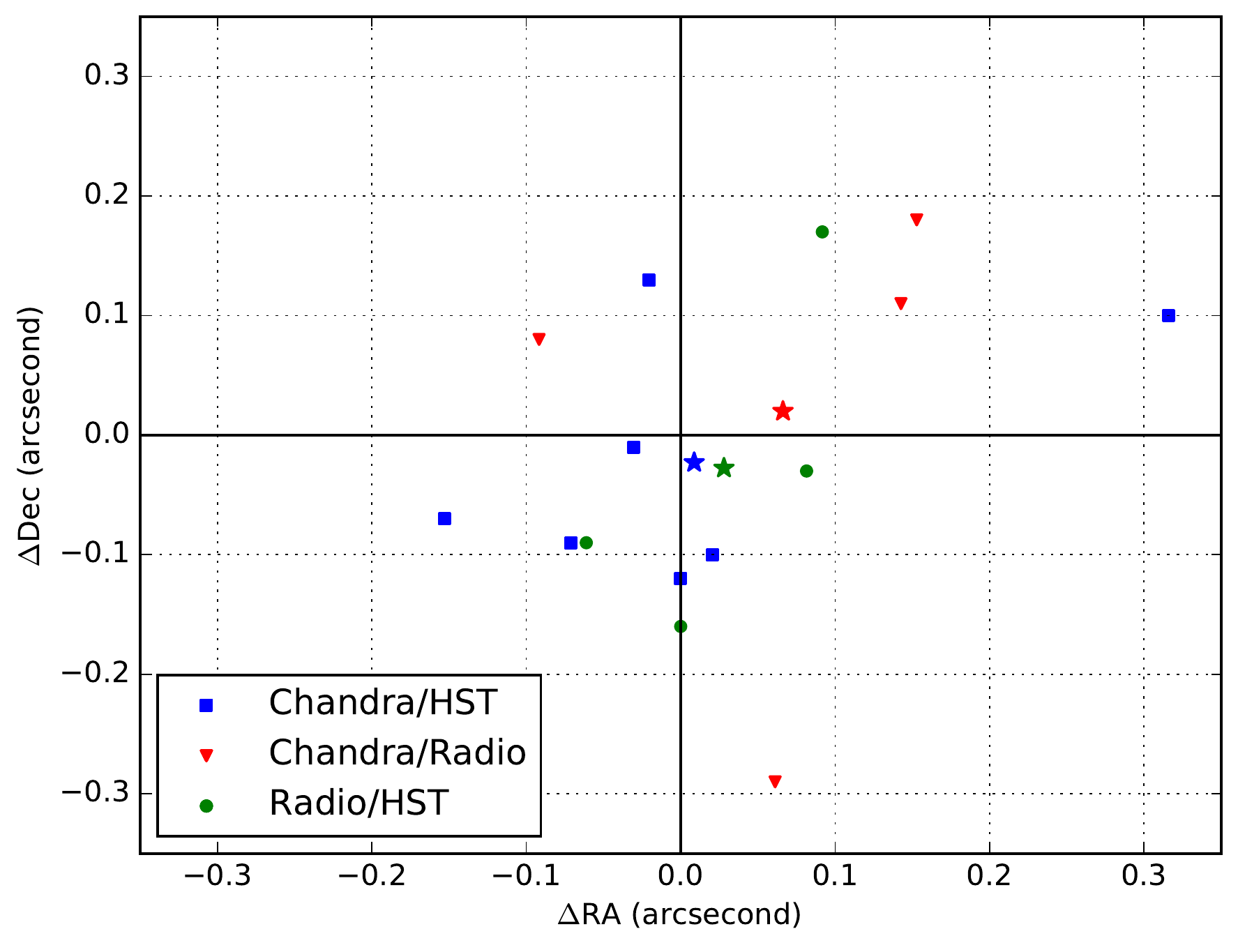}
 \caption{Relative positions of coincident, point-like {\it Chandra}, {\it HST} and VLA/Effelsberg sources. {\it Chandra} and {\it HST} coincidences are plotted as blue squares; {\it Chandra} and radio coincidences as red triangles; {\it HST} and radio coincidences as green circles. Stars represent the average offset between each pair of bands; systematic offsets are $<$0$^{\prime\prime}$.1 between any pair of bands. The residual discrepancies ($<$0$^{\prime\prime}$.4) are due to random scatter, mostly because of poor centroid determination of X-ray sources located far from the ACIS aimpoint.}
  \label{astrom_im}
  \vspace{0.3cm}
\end{figure}

 The positions of the X-ray sources were determined with the {\sc ciao} task \texttt{wavdetect}; the optical centroids of point-like sources with profile fitting in the {\sc ds9} package \citep{2003ASPC..295..489J}; the radio positions with the {\sc aips} package \texttt{imfit}. We found seven coincidences between X-ray and optical point-like sources, four coincident optical and radio sources, and four X-ray/radio coincidences. From a comparison of the optical/X-ray and optical/radio coincidences, we noticed that the {\it HST} astrometry was systematically offset by $\approx$0$^{\prime\prime}$.1 in both RA and Dec; therefore, we corrected the {\it HST} astrometry by that amount. The revised {\it HST} astrometry is still within the 0$^{\prime\prime}$.3 uncertainty of the 2MASS catalog. The X-ray positions (based on only 4 coincidences) were already within 0$^{\prime\prime}$.1 of the radio position and also within 0$^{\prime\prime}$.02 of the average of the improved {\it HST} positions. The resulting relative astrometry can be seen in Figure \ref{astrom_im}, with the average offsets between the three bands marked by stars. We did not find it necessary to make any further systematic corrections to any band. The residual discrepancies between the bands are simply random scatter of the individual sources. This is mostly due to the fact that the point spread function of point-like sources a few arcmin away from the {\it Chandra}/ACIS aimpoint becomes very extended and elongated, which increases the uncertainty in their central positions. In summary, we adopt the following X-ray positions for our ULXs: RA (J2000) = 13$^{\rm h}$29$^{\rm m}$39$^{\rm s}.960$, Dec. (J2000) = +47$^{\circ}$12$^{\prime}$36$^{\prime\prime}$.86 for ULX-1; RA (J2000) = 13$^{\rm h}$29$^{\rm m}$39$^{\rm s}.444$, Dec. (J2000) = +47$^{\circ}$12$^{\prime}$43$^{\prime\prime}$.60 for ULX-2. For both positions, we estimate a 3$\sigma$ error radius of 0$^{\prime\prime}$.4 arcsec, due to their off-axis positions and elongated PSF shape.


\section{Main Results}

\begin{table}
    \centering
    \caption{De-reddened brightness of the candidate optical counterparts. Brightness values have been corrected for a line-of-sight Galactic reddening $E(B-V) = 0.031$ mag, corresponding to an extinction $A_{\rm F435W} = 0.13$ mag, $A_{\rm F555W} = 0.10$ mag and $A_{\rm F814W} = 0.06$ mag. All three optical sources are consistent with bright giants (luminosity class II) or supergiants (class Ib), of approximate spectral type B5--B6 for the ULX-1 star, and B3--B4 for the ULX-2 stars.}
    \begin{tabular}{lrrr}
        \hline\hline
        \multicolumn{1}{c}{Source ID} & \multicolumn{1}{c}{F435W} & \multicolumn{1}{c}{F555W} & \multicolumn{1}{c}{F814W} \\
        & \multicolumn{1}{c}{(mag)} & \multicolumn{1}{c}{(mag)} & \multicolumn{1}{c}{(mag)} \\
        \hline
        ULX-1 & $24.62 \pm 0.03$ & $24.64 \pm 0.05$ & $24.87\pm0.07$ \\
        ULX-2 (north) & $23.14\pm0.03$ & $23.33\pm0.03$ & $23.59\pm0.03$ \\
        ULX-2 (south) & $23.60\pm0.04$ & $23.77\pm0.02$ & $23.87\pm0.04$ \\
        \hline\hline
    \end{tabular}
    \label{mags_tab}
\end{table}

\subsection{Point-like optical counterparts}

For ULX-1, only one bright optical candidate lies within the {\it Chandra} error circle (Figure \ref{field_im}, bottom left). This candidate is embedded in the brightest part of the elongated H$\alpha$ nebula (Figure \ref{LBT_slit_im}). For ULX-2, there are two potential optical counterparts within the {\it Chandra} error circle (Figure \ref{field_im}, bottom right), separated by $\approx$0$^{\prime\prime}$.25, with comparable brightness and colours. The relatively low precision of the {\it Chandra} position makes it impossible to rule out either candidate at this stage. The Vegamag brightnesses of the three sources in the {\it HST} bands (de-reddened for a line-of-sight extinction corresponding to $A_V = 0.095$ mag, with Cardelli extinction law) are summarized in Table \ref{mags_tab}. At a distance modulus of 29.67 mag, the absolute magnitudes and colours of all three sources are consistent with those of bright giants (luminosity class II) or blue supergiants (class Ib); if two of those three sources are the true counterparts of ULX-1 and ULX-2, they are likely to contain also a contribution from disk emission. For simplicity, we investigate two extreme scenarios; first, purely emission from the stellar companion and second, purely emission from the irradiated disk. It is likely that the true emission is some combination of both processes, and thus we investigate the upper limits of stellar brightness and irradiation in this section.
 
\begin{figure*}
\centering
\includegraphics[width=0.48\textwidth]{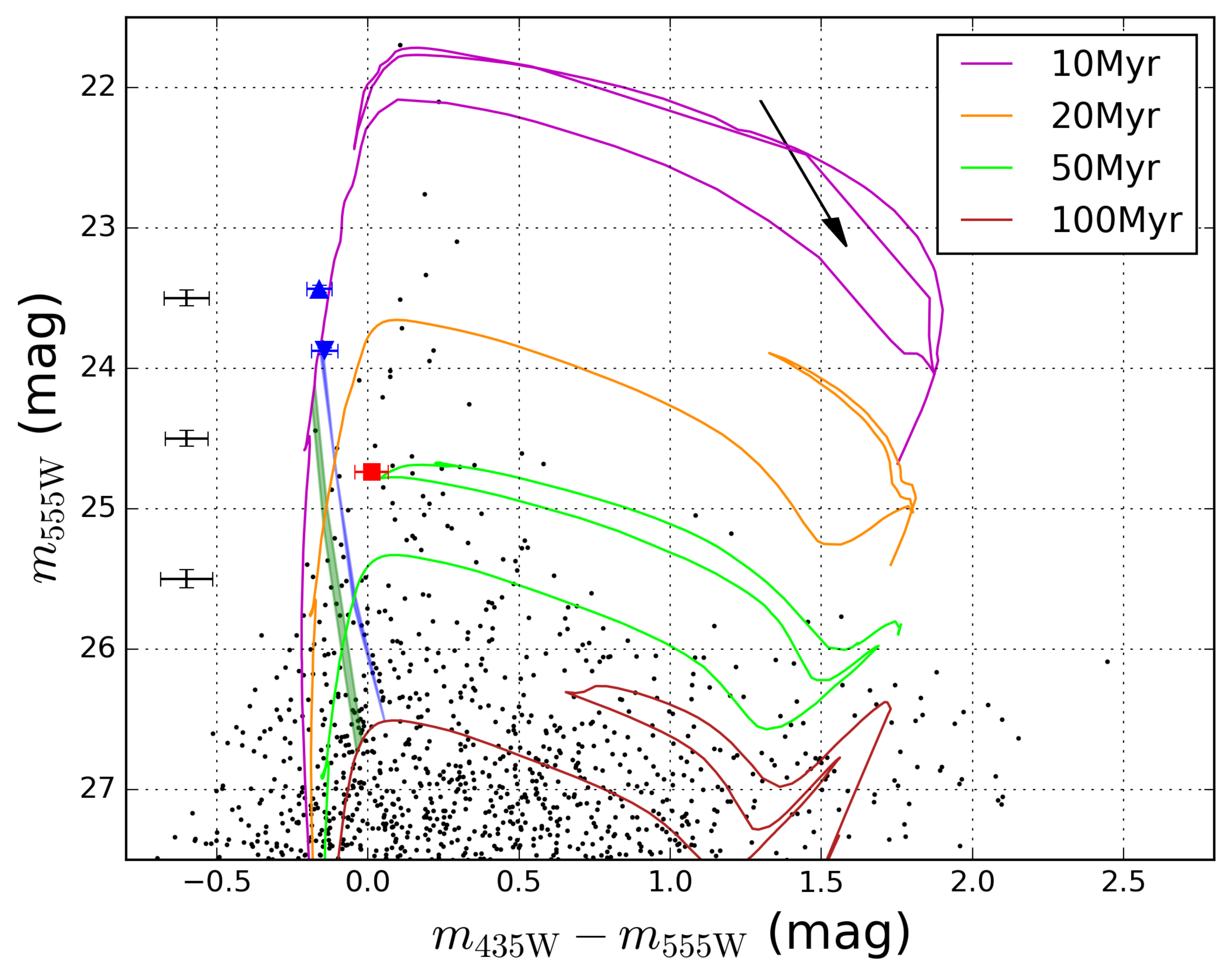}
\includegraphics[width=0.48\textwidth]{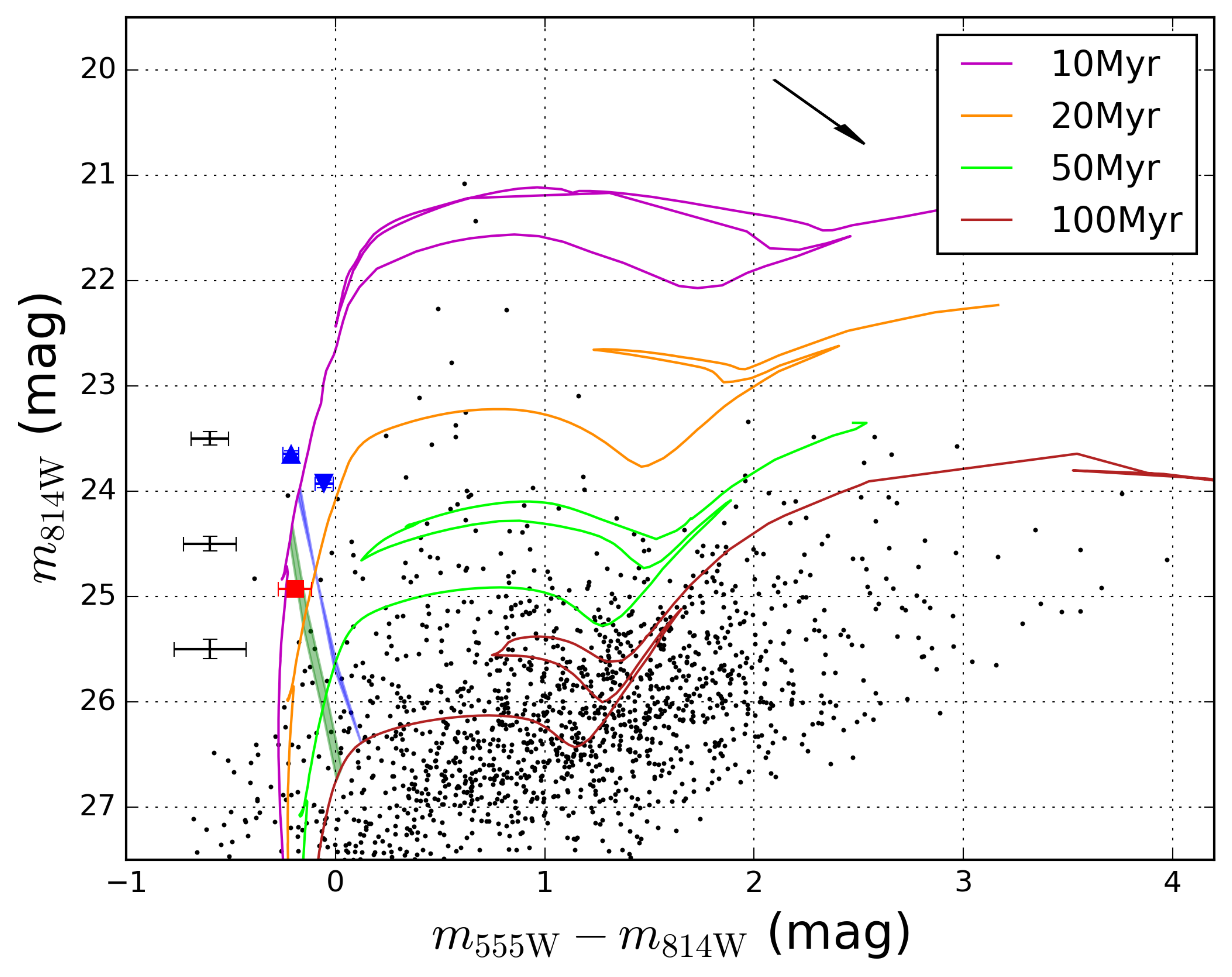}\\
\includegraphics[width=0.48\textwidth]{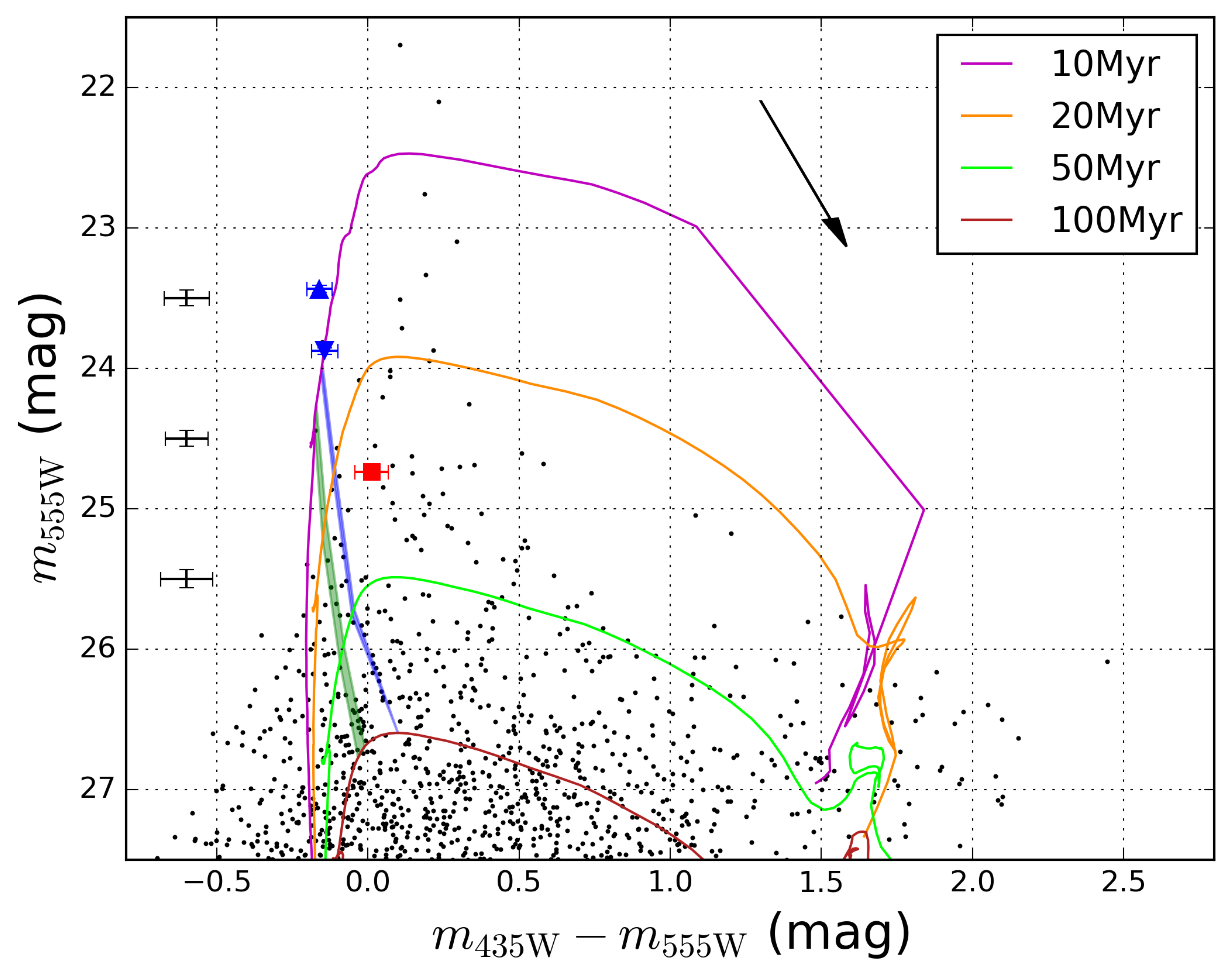}
\includegraphics[width=0.48\textwidth]{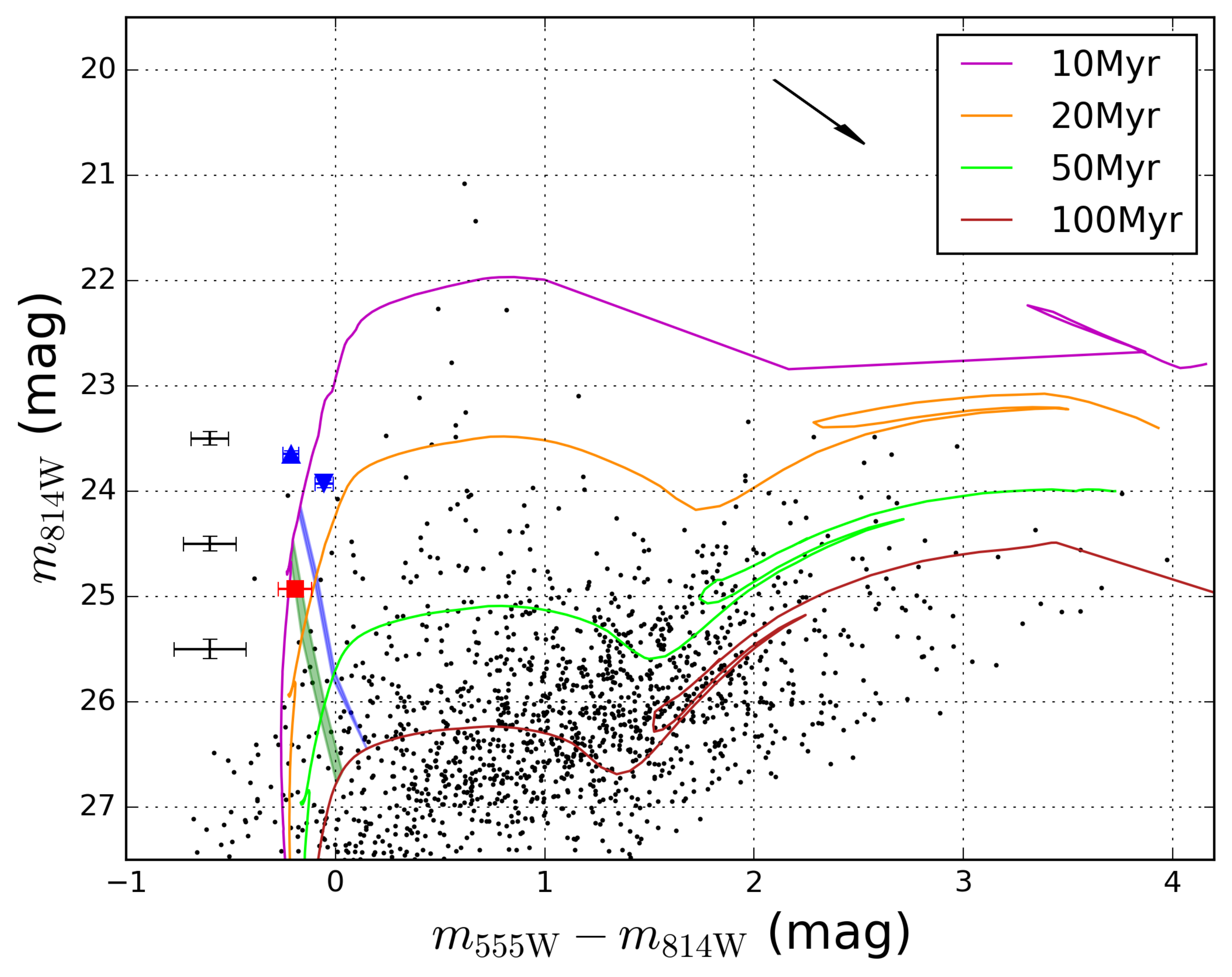}
 \caption{Top left panel: colour-magnitude diagram in the (F435W $-$ F555W) versus F555W bands. The red square represents the candidate optical counterpart of ULX-1. The blue triangle represents the northern candidate counterpart for ULX-2 while the upside-down blue triangle represents the southern candidate counterpart. Black datapoints show the surrounding stars, with the black error bars on the left indicating their average uncertainties for each F555W magnitude interval. The black arrow shows the effect of one magnitude of extinction in F555W. Overlaid are Padova stellar isochrones for metallicity $Z = 0.015$. The isochrones have been corrected for a line-of-sight extinction $A_V = 0.095$ mag. The green and blue vertical bands drawn across the isochrones represent the expected brightness and colour of Roche-lobe-filling stars with stellar densities consistent with a binary period of 6 or 12 days respectively, as seen in the X-ray light-curve of ULX-1 (see Section 3.1). Top right panel: as in the top left panel, but for F555W $-$ F814W versus F814W. Bottom left panel: as in the top left panel, but with Padova stellar isochrones corresponding to a metallicity $Z = 0.040$. Bottom right panel: as for the top right panel, but for $Z = 0.040$.}
  \label{colmag_im}
  \vspace{0.3cm}
\end{figure*}

To constrain the ages of the three sources and of the surrounding stellar population, we used colour-magnitude diagrams in the three {\it HST} broadband filters (Figure \ref{colmag_im}); we plotted all stars with reliable brightness values within the 15$^{\prime\prime}\times$15$^{\prime\prime}$ field shown in Figure \ref{field_im}. The black arrow in each plot indicates the effect of a hypothetical additional intrinsic extinction $A_V = 1$ mag. We compared the data with a set of theoretical isochrones \footnote{Available at http://stev.oapd.inaf.it/cgi-bin/cmd} \citep{2012MNRAS.427..127B,2015MNRAS.452.1068C} for a metallicity $Z = 0.015$ (Figure \ref{colmag_im}, top panels) and $Z=0.040$ (Figure \ref{colmag_im}, bottom panels). The choice of those two alternative metal abundances reflects the two discrepant classes of values generally quoted in the literature for the disk of M\,51; see, {\it e.g.},  \cite{2015ApJ...808...42C} and \cite{2004ApJ...615..228B} for the argument in favour of slightly sub-solar metallicities, and \cite{2010ApJS..190..233M} and \cite{1994ApJ...420...87Z} in favour of super-solar values. We will return to this metallicity discrepancy in Section 3.2.2.

Comparing the colour-magnitude plots, we notice that the inferred ages of the optical companion of ULX-1 are not consistent: $\approx$30 Myr in the (F435W $-−$ F555W) versus F555W diagram, and $\approx$10--20 Myr in the (F555W $−-$ F814W) versus F814W diagram, regardless of metal abundance. The two ages cannot be made consistent with the addition of intrinsic extinction; the star is $\approx$0.15 mag brighter in the F555W band than predicted by stellar isochrones. The reason for this small discrepancy is unclear. We note that the source is embedded in the most luminous part of the optical nebula, with its strong [O {\sc iii}] $\lambda$4959,5007 emission lines falling into the F555W band, which makes the subtraction of the local background emission more challenging and may be responsible for the small excess in that band. Alternatively, some of the optical emission from that point-like source is coming from the accretion disk or the outflow around the ULX, so that its colours are slightly different from a pure stellar spectrum. Distinguishing between the optical emission from donor star and accretion flow is a notoriously difficult task in ULXs \citep{2014MNRAS.442.1054H, 2013ApJS..206...14G, 2012ApJ...750..152S, 2012ApJ...745..123G, 2012ApJ...750..110T, 2011ApJ...737...81T}; we will return to this issue later in this Section.

Keeping those caveats in mind, we conclude that the optical counterpart for ULX-1 has a most likely age of $\sim$20 Myr, if dominated by a single donor star. We also plot stellar density bands (blue and green shaded regions in Figure \ref{colmag_im}). These shaded regions indicate regions of the parameter space where stars have a stellar density consistent with a Roche-lobe-overflow binary of period 6 or 12 days respectively. These two alternative values of the period come from the recurrent eclipses found in the X-ray light-curves for ULX-1 \citep{2016ApJ...831...56U}; the 6-day period is a solution suitable for mass ratios $4 \la q \la 10$, while the 12-day period is the solution corresponding to mass ratios $0.5 \la q \la 1$, where $q \equiv M_2/M_1$. Those two characteristic periods are then converted to stellar densities via the period-density-$q$ relation \citep{1983MNRAS.204..449E}. The current optical data (taking into account also the possible contamination from disk and nebular emission) are not sufficient to test whether the candidate star sits in either band. Thus we cannot yet use this method to rule out either of the proposed orbital periods of ULX-1, and constrain its mass ratio. However, we illustrate this method as a proof of concept, and to highlight the fact that the candidate optical star is at least approximately consistent with the kind of Roche-lobe-filling donor star required for ULX-1.

\begin{table}

\centering
\caption{Extinction-corrected emission line fluxes for the nebulae around ULX-1 and ULX-2, inferred from our LBT spectra and scaled by the {\it HST} images. Fluxes were corrected for a line-of-sight extinction $A_V = 0.095$ mag. All fluxes were re-normalized so that the H$\alpha$ $+$ [N {\sc ii}] fluxes match the corresponding line fluxes measured from the {\it HST} image. Fluxes are displayed in terms of their strength relative to the H$\beta$ line ($100F_{\lambda}/F_{\rm H\beta}$).}
\label{ulx_tab}
\begin{tabular}{lrr}
\hline\hline
\multicolumn{1}{c}{Line} & \multicolumn{2}{c}{$100\times(f_{\lambda}/f_{\rm H\beta})$} \\
 & \multicolumn{1}{c}{ULX-1} & \multicolumn{1}{c}{ULX-2} \\
\hline\hline
3426 {[}Ne {\sc v}{]} & \Pl{16}{1} & \\
3728 {[}O {\sc ii}{]} ($\lambda 3726 + \lambda 3729$) & \Pl{605}{28} & \Pl{88}{6}\\
\hspace{0.3cm}3726 {[}O {\sc ii}{]}$^a$ & \Pl{243}{11} &\\
\hspace{0.3cm}3729 {[}O {\sc ii}{]}$^a$ & \Pl{353}{16} &\\
3869 {[}Ne {\sc iii}{]} & \Pl{92}{5} & \\
3889 H$\zeta$ $+$ He {\sc i} & \Pl{17}{2} & \\
3967 {[}Ne {\sc iii}{]} $+$ H$\epsilon\,\lambda 3970$   & \Pl{47}{2} & \\
4069 {[}S {\sc ii}{]}$^b$  & \Pl{22}{5} & \\
4104 H$\delta$ & \Pl{24}{2} & \\
4340 H$\gamma$  & \Pl{46}{3} & \Pl{30}{2}\\
4363 {[}O {\sc iii}{]} & \Pl{23}{2} & \\
4686 He {\sc ii} & \Pl{22}{1} &\\
4861 H$\beta$ & 100 & 100 \\
4959 {[}O {\sc iii}{]}  & \Pl{207}{10} & \Pl{30}{3} \\
5007 {[}O {\sc iii}{]} & \Pl{604}{28} & \Pl{96}{5} \\
5200 {[}N {\sc i}{]} ($\lambda 5198 + \lambda 5202$) & \Pl{15}{1} & \\
6300 {[}O {\sc i}{]}  & \Pl{66}{3} & \\
6364 {[}O {\sc i}{]}  & \Pl{24}{2} & \\
6548 {[}N {\sc ii}{]} & \Pl{127}{6} & \Pl{48}{2} \\
6563 H$\alpha$ & \Pl{299}{14} & \Pl{307}{15} \\
6583 {[}N {\sc ii}{]} & \Pl{377}{19} & \Pl{95}{5} \\
6716 {[}S {\sc ii}{]}  & \Pl{132}{6} & \Pl{51}{3} \\
6731 {[}S {\sc ii}{]}  & \Pl{111}{5} & \Pl{35}{3} \\
7136 {[}Ar {\sc iii}{]} & \Pl{21}{1} & \\
7320 {[}O {\sc ii}{]}  & \Pl{51}{4} &\\
9069 {[}S {\sc iii}{]}  & \Pl{20}{1} &\\
9532 {[}S {\sc iii}{]}  & \Pl{40}{3} &\\
10049 H {\sc i} Pa$\alpha$ & \Pl{3.0}{0.2} &\\
12818 H {\sc i} Pa$\beta$ & \Pl{2.2}{0.3} &\\
*H$\beta$ flux ($10^{-16}\,\flux$) & \Pl{4.21}{0.18} & \Pl{4.47}{0.17}\\
\hline
\end{tabular}
    \begin{flushleft}
    $^a$ Deblended with {\sc iraf} line-fitting tools\\
    $^b$ Slightly contaminated by {[}S {\sc ii}{]} $\lambda 4076$\\
    \end{flushleft}
\end{table}

For ULX-2, both candidate optical counterparts have ages consistent with $\approx$10 Myr, and mass $M \approx 18 M_{\odot}$, if we correct for Galactic line-of-sight extinction only. If we also include additional dust reddening within M\,51 equal to the Galactic value, the inferred age of both stars is instead $\approx$8 Myr, and their masses $M \approx 22$--23 $M_{\odot}$. Stars at this evolutionary stage have a characteristic radius $\sim$ 30 $R_{\odot}$ and a characteristic density $\rho \sim 3\E{-4}$ g cm$^{-3}$. This would imply a characteristic binary period of $24$ days for $q \approx 1$ ({\it i.e.} for the case of a canonical black hole primary), or a period of $10$ days for $q \approx 17$ ({\it i.e.} for the case of a canonical neutron star primary). The period of ULX-2 could not be directly measured from its X-ray light-curve \citep{2016ApJ...831...56U} and so we do not have enough observational constraint to test this prediction yet; however, there is at least a plausible hint, from the average eclipse fraction, that the period is $\sim$10 days \citep{2016ApJ...831...56U}, consistent with the neutron star scenario.

From Figure \ref{colmag_im}, we can see that both potential ULX-2 optical counterparts appear to be outliers, much brighter and bluer than the surrounding stars. We considered and dismissed the possibility that they are the counterparts of two luminous X-ray sources, unresolved in the {\it Chandra} image. The reason this is unlikely is that ULX-2 shows sharp, deep eclipses, during which the residual luminosity is $L\approx3\times10^{37}\,\ergs$. Barring implausible coincidences, this must be the upper limit to the luminosity of any potential second X-ray source. Such a luminosity would not be enough to contribute significantly to the optical emission, via reprocessing of X-ray irradiation, given that both optical sources have $M_{\rm V} \sim -6$ mag. Thus, we conclude that at least one of the two blue point-like sources in the {\it Chandra} error circle for ULX-2 is an intrinsically bright star, while the other source may either be dominated by irradiation, or also be an intrinsically bright donor star.

\begin{figure}
\centering
\includegraphics[width=0.49\textwidth]{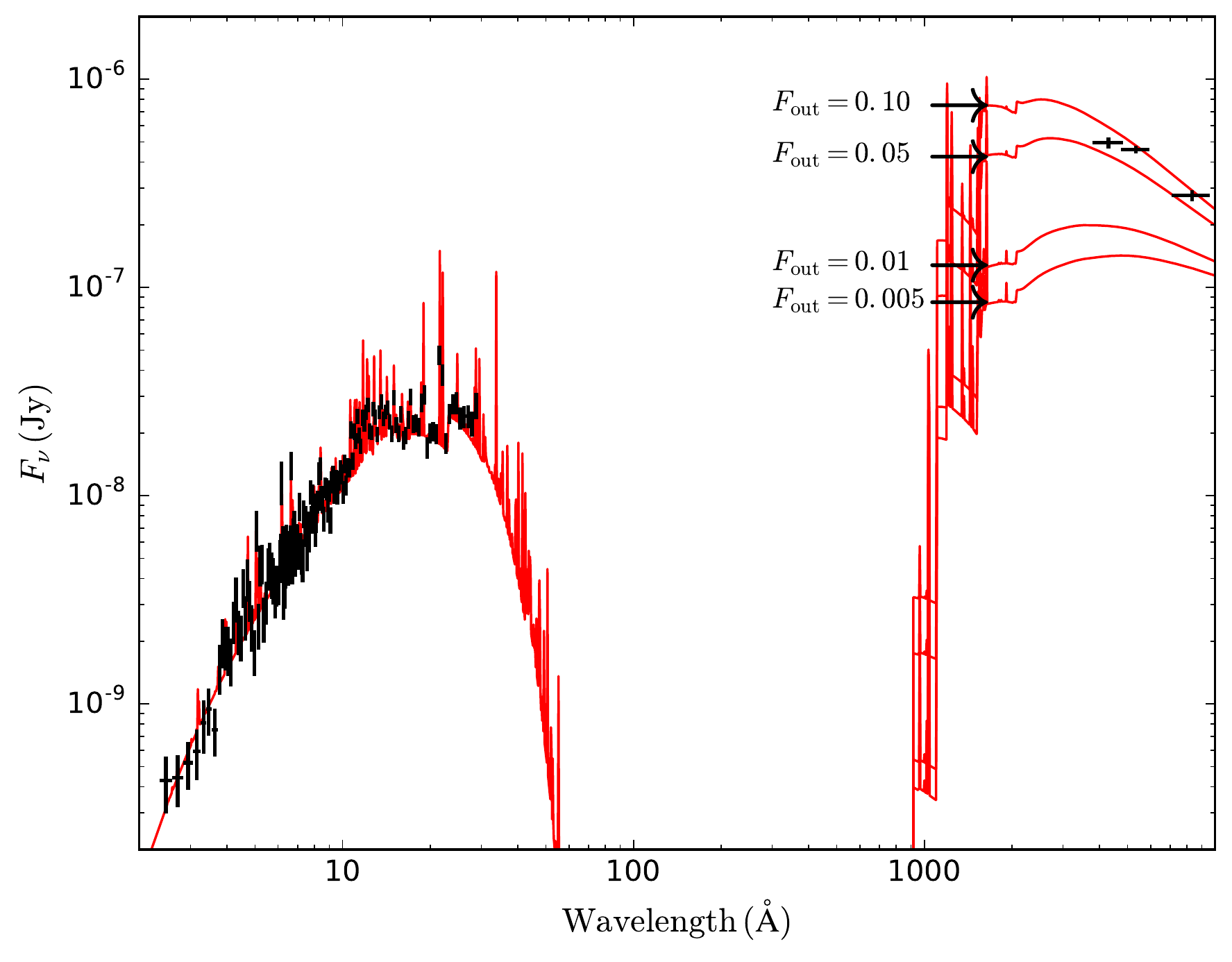}
\vspace{-0.4cm}
 \caption{Stacked {\it Chandra}/ACIS-S spectrum and {\it HST} datapoints (black) for ULX-1, fitted with a {\it diskir} plus thermal plasma models (red curves). A sequence of four models has been plotted, differing only for their optical reprocessing fraction $f_{\rm out}$. A very high reprocessing fraction (almost 10\%) is formally required in this model to reproduce the high optical/X-ray flux ratio. Such a high value strongly suggests that we are seeing a significant contribution to the optical continuum flux from the donor star, although we cannot separate the disk and star components with the data at hand.}
  \label{ulx1_xray_spec_im}
  \vspace{0.3cm}
\end{figure}

Next, we try to quantify whether X-ray irradiation can significantly contribute to the optical counterpart of ULX-1 and to one of the two blue stars at the location of ULX-2. The X-ray spectrum of ULX-1 is well fitted by an irradiated disk model \citep{2016ApJ...831...56U}, for example the {\sc XSPEC} model {\it diskir} \citep{2009MNRAS.392.1106G}. In Figure \ref{ulx1_xray_spec_im}, we plot the combined X-ray spectrum and optical data points, together with a sequence of {\it diskir} models with varying levels of reprocessing ($f_{\rm out} = 0.1$, 0.05, 0.01 and 0.005). If we want to explain the optical flux as the intercepted and reprocessed component of the X-ray flux, we need a reprocessing fraction $f_{\rm out} \approx 0.075$. This is an order of magnitude higher than predicted \citep{1999MNRAS.303..139D, 1997ApJ...488...89K,1996A&A...314..484D,1990A&A...235..162V} or observed \citep{2014MNRAS.439.1390R,2009MNRAS.392.1106G,2002MNRAS.331..169H} in Galactic X-ray binaries with sub-Eddington standard disks, although \citet{2010MNRAS.407.2166G} found that the reprocessing fraction could be as high as 20\% in the Galactic black hole binary GX 339-4. However, much less is known about the reprocessing fraction in ULXs. For a transient ULX in M\,83, \citet{2012ApJ...750..152S} estimated a reprocessing fraction of $f \approx 5 \times 10^{-3}$; for other ULXs, reprocessing fractions of a few $10^{-2}$ were inferred \citep{2014MNRAS.444.2415S}. Super-critical accreting sources have geometrically thicker disks in the inner region, but even so, the disk alone cannot directly intercept such a high fraction of the X-ray emission; however, if the ULX has a broad disk outflow and polar funnel geometry, some of the X-ray photons emitted along the funnel may be scattered by the wind and irradiate the outer disk \citep{2014MNRAS.444.2415S, 2017MNRAS.469.2997N}. Indeed, we argued \citep{2016ApJ...831...56U} that ULX-1 may have a strong wind, because of its soft X-ray spectral residuals, which in other ULXs are associated with outflows \citep{2015MNRAS.454.3134M, 2016Natur.533...64P}. Nonetheless, we also know that ULX-1 must have a fairly large donor star, because of its eclipses, and we have already shown that the inferred size of the star is roughly consistent with the observed optical flux. Alternatively, the non-simultaneity of the X-ray and optical data may contribute to the unphysically high reprocessing fraction that is required. However, studying the long- and short-term X-ray variability of ULX-1 (and ULX-2) from {\it Chandra}, {\it XMM-Newton} and {\it Swift}\footnote{For {\it XMM-Newton} and {\it Swift}, ULX-1 and ULX-2 are unresolved and thus we can only study their combined variability.}, we find that the source only varies by a factor of $\approx2$. Any change in the optical flux (assuming irradiation of the disk) must be less than that amount, because the optical emission comes from the Rayleigh-Jeans tail of the disk spectrum and it includes an intrinsic non-variable contribution from the stellar companion.
We can use a similar argument for ULX-2. In that case, the best-fitting {\sc xspec} model to the X-ray spectrum reported by \citep{2016ApJ...831...56U} is a {\it diskpbb} model (slim disk) rather than {\it diskir}, so we cannot directly constrain the reprocessing fraction from the model parameters. However, forcing a fit of the same spectrum with {\it diskir} we obtain a reprocessing fraction $\approx 0.2$; this is because the observed X-ray flux is approximately the same as from ULX-1, while the optical flux is three times brighter. That makes it even more difficult to explain the optical emission as dominated by an irradiated disk rather than by a supergiant donor star.

\subsection{Ionized nebulae} \label{lbt_results_sec}

\begin{figure*}
\centering
\includegraphics[width=0.49\textwidth]{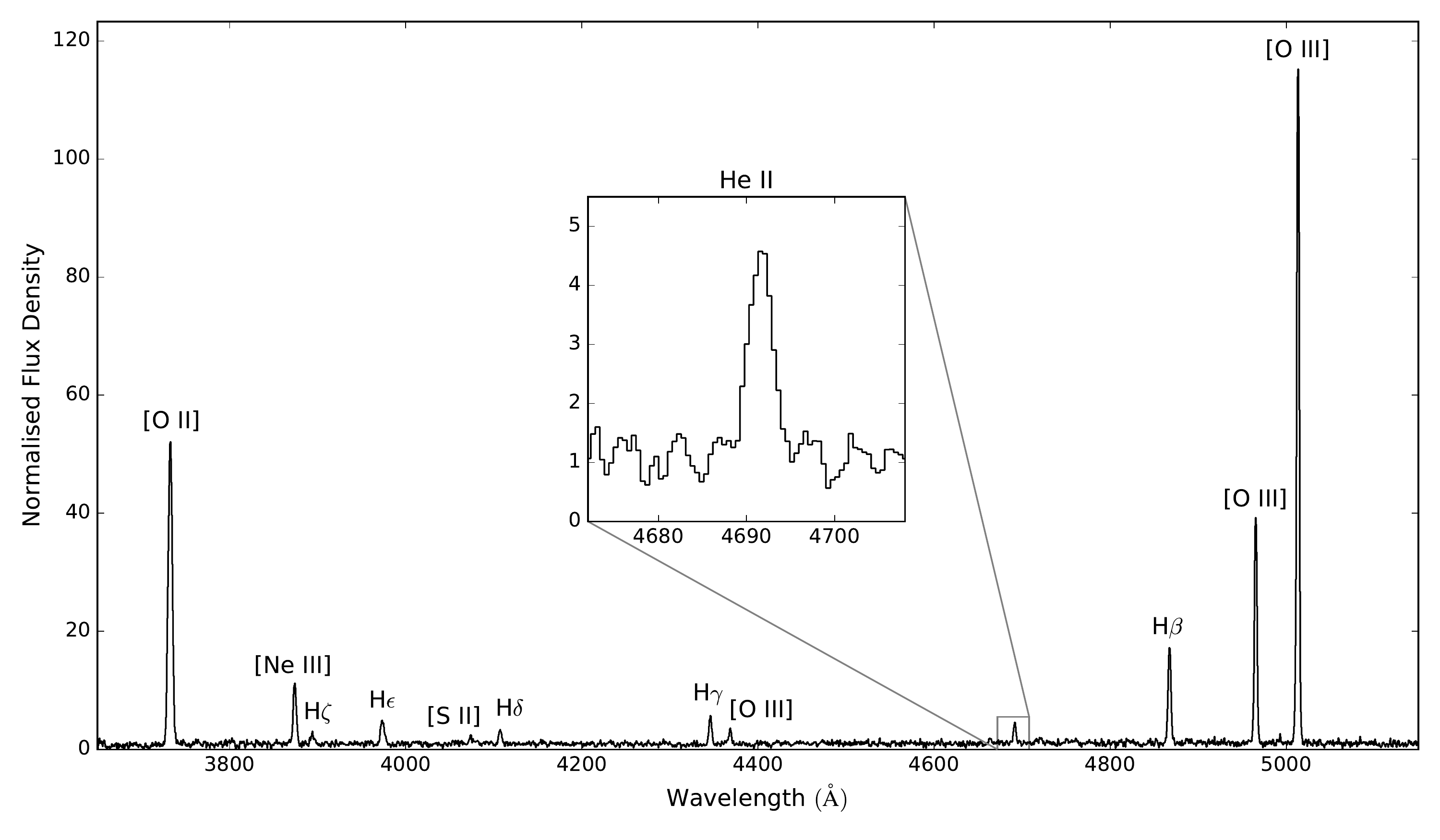}
\includegraphics[width=0.49\textwidth]{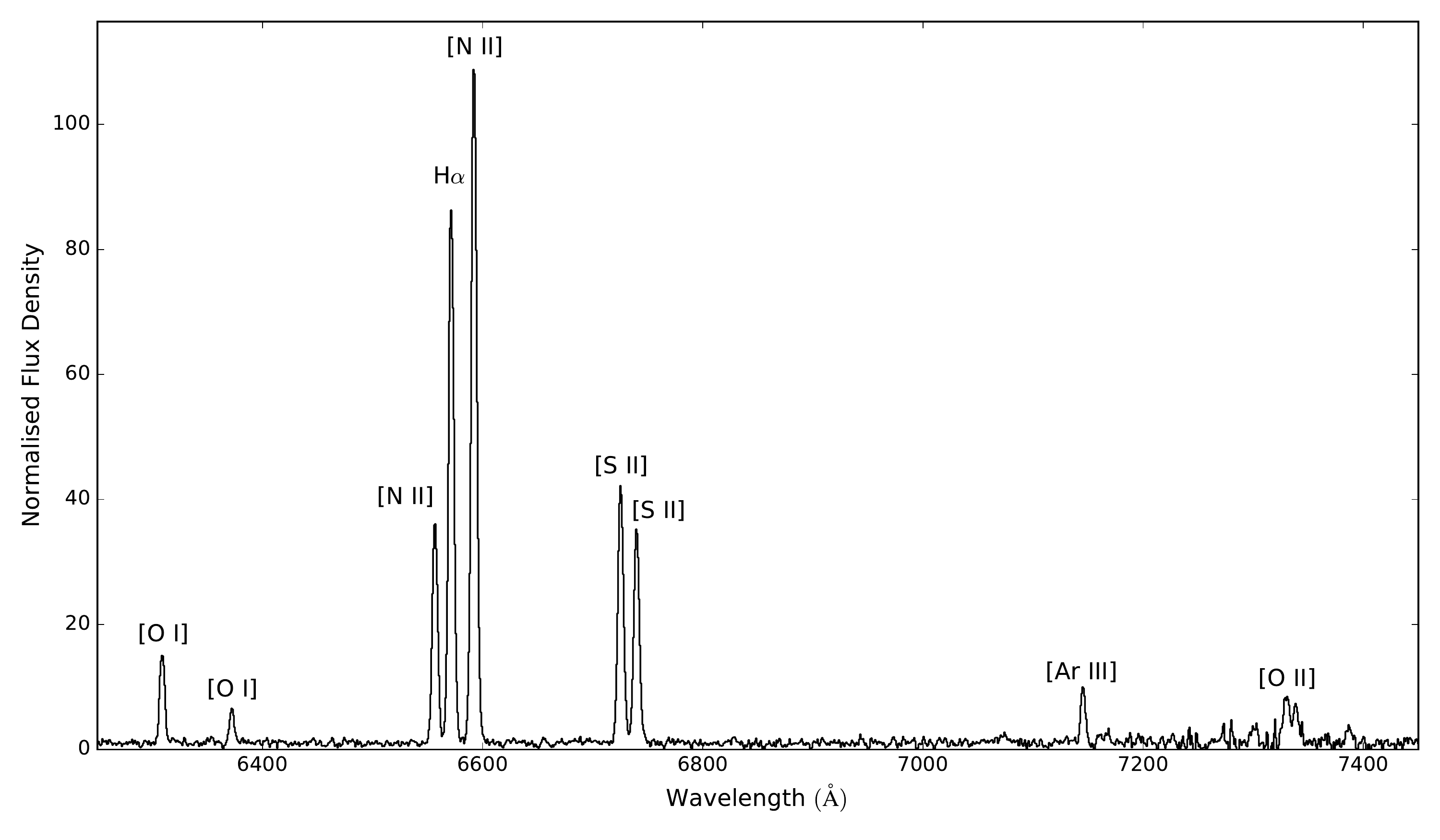}
 \caption{Left panel: LBT MODS1 plus MODS2 spectrum (blue arm) of the nebula around ULX-1, normalized to the continuum. The spectrum comes from the OB1 slit configuration (PA $= 80^{\circ}$), which is along the long axis of the jet-like nebula.  Inset: zoomed-in view of the He {\sc ii} $\lambda$4686 emission line. Right panel: as in left panel, but for the red side of the spectrum. (Note the different flux density scales of the two panels.)}
  \label{ulx1_opt_lines_im}
  \vspace{0.3cm}
\end{figure*}

\begin{figure*}
\centering
\includegraphics[width=0.49\textwidth]{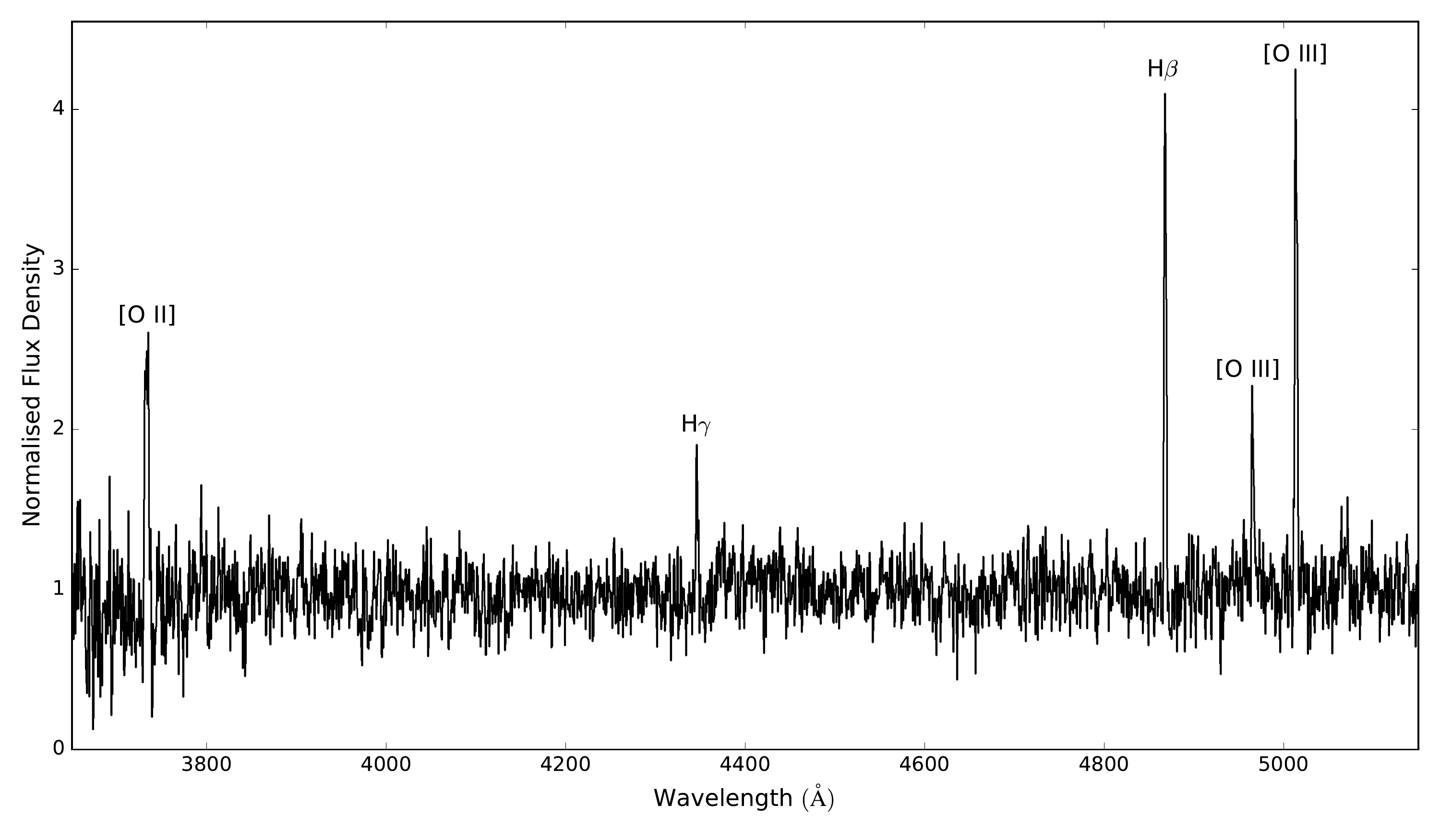}
\includegraphics[width=0.49\textwidth]{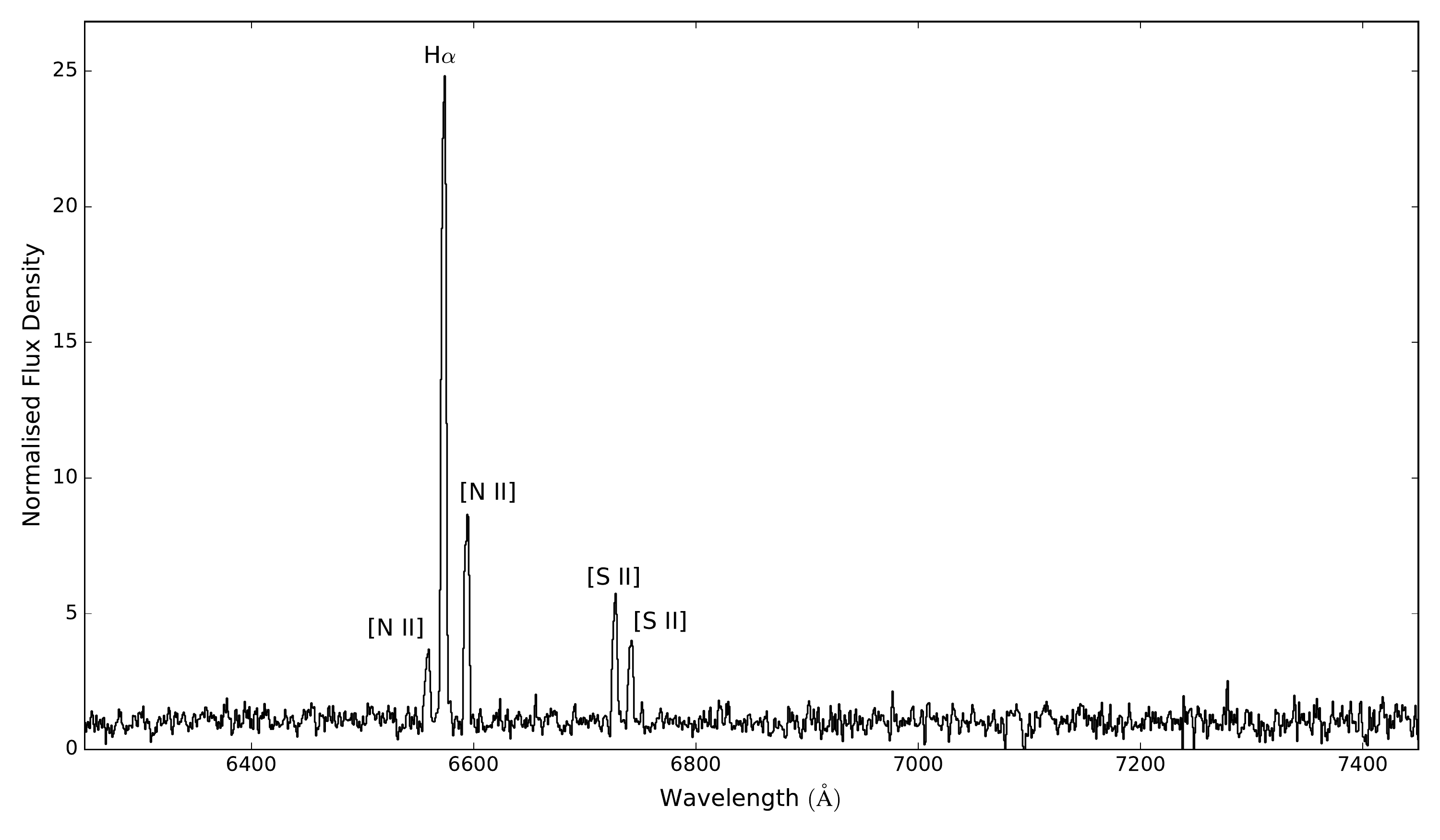}
 \caption{Left panel: LBT MODS1 plus MODS2 spectrum (blue arm) of the nebula around ULX-2, normalized to the continuum. The spectrum comes from the OB2 slit configuration (PA $= 322^{\circ}$.5); the stellar continuum includes both candidate optical counterparts of ULX-2.  Right panel: as in the left panel, but for red side of the spectrum. (Note the different flux density scales of the two panels.)}
  \label{ulx2_opt_lines_im}
  \vspace{0.3cm}
\end{figure*}


The {\it HST}/ACS image in the continuum-subtracted F658N filter clearly shows the presence of line emission around both ULXs (Figures \ref{field_im} and \ref{LBT_slit_im}). The nebula around ULX-1 is elongated, with a length of $\approx$2$^{\prime\prime}$.5 $\approx$100 pc and a width of $\approx$0$^{\prime\prime}$.7 $\approx$30 pc. The nebula around ULX-2 is more circular, with an outer radius of $\approx$1$^{\prime\prime}$.6 $\approx$65 pc. There is also another region of line emission $\approx$200 pc to the northwest of ULX-2, of similar size and brightness, but without any bright X-ray source inside. In the absence of other evidence, we assume that this nebula is unrelated to ULX-2, rather than (for example) being part of a double-lobed structure powered by the ULX.

We defined source and background regions suitable for the two nebulae, and extracted their net count rates. We obtain rates of $21.52 \pm 0.10$ ct s$^{-1}$ and $12.79 \pm 0.25$ ct s$^{-1}$ for the ULX-1 and ULX-2 nebulae, respectively, in the F658N band. We then used the {\it HST}/ACS Zeropoint tables available online\footnote{https://acszeropoints.stsci.edu/} (see also \citealt{2016AJ....152...60B}) to convert count rates to fluxes. A count rate of 1 ct s$^{-1}$ corresponds to a flux of $\approx$1.95 $\times 10^{-18}$ erg cm$^{-2}$ s$^{-1}$ \AA$^{-1}$ in that band. The effective width of the filter is 75 \AA. Thus, the integrated line fluxes from the two nebulae are $(3.39 \pm  0.02) \times 10^{-15}$ erg cm$^{-2}$ s$^{-1}$ and $(2.01 \pm  0.04) \times 10^{-15}$ erg cm$^{-2}$ s$^{-1}$, respectively, where we have also applied a de-reddening correction for line-of-sight extinction. At a distance of 8.6 Mpc, this corresponds to line luminosities of $\approx$3.0 $\times 10^{37}$ erg s$^{-1}$ and $\approx$1.8 $\times 10^{37}$ erg s$^{-1}$, respectively, in the F658N filter. The filter bandpass covers [N {\sc ii}] $\lambda 6548$, H$\alpha$ $\lambda 6563$, and [N {\sc ii}] $\lambda 6583$. Those three lines are redshifted by $\Delta \lambda \approx 9$ \AA\ with respect to their rest-frame wavelengths, because of the systemic velocity of M\,51. This ensures that the flux from the [N {\sc ii}] $\lambda 6548$ line is entirely included in the F658N bandpass, given the moderate line widths that we will discuss later in this Section. Of course the {\it HST} image alone cannot tell us how the line flux is split between H$\alpha$ and the [N {\sc ii}] lines. For this, we need to examine the LBT spectra.

The LBT spectra across the ULX-1 and ULX-2 nebulae are displayed in Figures \ref{ulx1_opt_lines_im} and \ref{ulx2_opt_lines_im} respectively. The emission lines significantly detected in the two spectra, and their flux ratios to H$\beta$, are listed in Table \ref{ulx_tab}. Those values are the total emission from the whole of each nebula. We recall that we obtain the total flux of each line by rescaling the H$\alpha$ $+$ [N {\sc ii}] flux measured in the LBT spectrum to the total flux measured from the {\it HST} image in the F658N filter, and then rescaling the LBT fluxes of every other line according to their ratios to H$\alpha$.  We also recall that ULX-1 was observed both in the OB1 (slit along the bubble) and OB2 (slit across the bubble) configurations. To a first approximation, the two spectra are identical; in this paper, we use the spectra for the OB1 slit, because they provide a better approximation to the average emission of the whole nebula. A more comprehensive comparison of the two spectra and of the internal structure of the ULX-1 nebula is left to further work. 

From the line ratios (Table \ref{ulx_tab}), we find Balmer decrements  
H$\alpha$:H$\beta = 2.99 \pm 0.14$, H$\gamma$:H$\beta = 0.46 \pm 0.03$ in the ULX-1 nebula; in the ULX-2 nebula, 
H$\alpha$:H$\beta = 3.07 \pm 0.15$, H$\gamma$:H$\beta = 0.30 \pm 0.02$. The canonical values for Case B recombination are 2.87:1:0.47 at 10,000 K, and 2.75:1:0.48 at 20,000 K \citep{2006agna.book.....O}. However, the Balmer decrement is slightly steeper for radiative shocks: for solar metallicity and a shock velocity of 150 km s$^{-1}$, the ratios are 3.06:1:0.46, while for a shock velocity of 500 km s$^{-1}$, the Balmer decrement is 2.92:1:0.47 \citep{2008ApJS..178...20A}. Thus, the observed Balmer decrement in the ULX-1 nebula is perfectly consistent with shock-ionized gas. Instead, for the ULX-2 nebula, the observed decrement is steeper than expected. Part of the reason for the discrepancy may be that a few percent of the flux in the bluer part of the spectrum is lost (compared with the flux in the red spectrum) because of differential refraction, as we mentioned in Section 2.2; this affects particularly H$\gamma$ and particularly the ULX-2 spectrum because of the slit orientation in the OB2 configuration (Section 2.2). In addition, we have only corrected for line-of-sight extinction; if we add an intrinsic dust reddening comparable to the Galactic reddening (Section 2.2), the extinction-corrected H$\alpha$:H$\beta$ ratio is reduced by 4\% and the H$\gamma$:H$\beta$ ratio is increased by 3\%. Overall, those small uncertainties do not affect our main results and interpretation of the spectra. In the rest of this section, we focus on the most significant properties of the two nebulae, and their main differences.


\subsubsection{ULX-1 nebula}

We see strong [N {\sc ii}] and [S {\sc ii}] lines relative to H$\alpha$ (Figure \ref{ulx1_opt_lines_im}), with a flux ratio ([N {\sc ii}] $\lambda$6548 + [N {\sc ii}] $\lambda$6583)/H$\alpha$ $\approx$ 1.7 and ([S {\sc ii}] $\lambda$6716 + [S {\sc ii}] $\lambda$6731)/H$\alpha$ $\approx$ 0.81. Such high ratios (even after accounting for a possible super-solar abundance in the disk of M\,51)  are traditionally good indicators of shock-ionized material \citep{1973ApJ...180..725M,1997ApJS..108..261B, 1998ApJS..117...89G}. Based on those ratios alone, the bubble could be classified as a typical supernova remnant; however, its very elongated morphology and the presence of a ULX in the middle make it more likely that the gas is shocked by a jet emitted from the compact object, as is the case in other ULX bubbles. The ULX-1 bubble also shows He {\sc ii} $\lambda$4686 (Figure \ref{ulx1_opt_lines_im}) and [N {\sc v}] $\lambda$3426 (Table \ref{ulx_tab}) emission.
Both features are consistent with shock-ionized gas and/or X-ray photo-ionized gas, but are not consistent with stellar photo-ionization. 

In order to estimate the electron temperature and density of the gas, we used standard diagnostic line ratios \citep{1989agna.book.....O}, coded in the task \texttt{temden} in {\sc iraf} \citep{1994ASPC...61..327S}. For the [O {\sc iii}] line (sensitive to electron temperature), we measure an intensity ratio $I(\lambda 4959 + \lambda 5007)/I(\lambda 4363) \approx 35$, which corresponds to $T_e$[O {\sc iii}]$ = 22,000 \pm 2000$ K. From the [O {\sc ii}] line intensity ratio $I(\lambda 3726 + \lambda 3729)/I(\lambda 7320 + \lambda 7330) \approx 12$, we measure $T_e$[O {\sc ii}]$ = 20,000 \pm 1000$ K. (In general, those two temperature values do not need to agree, as they may be due to emission coming from different regions.) Those temperatures are a factor of two higher than expected for typical H {\sc ii} regions, and suggest that there is a significant component of shock-ionized gas. For the [S {\sc ii}] line (sensitive to electron density), we find an intensity ratio [S {\sc ii}] $I(\lambda 6716)/I(\lambda 6731) \approx 1.20$. Around the previously estimated values of $T_e$, this ratio corresponds to $n_e = (280 \pm 60)$ cm$^{-3}$. This is to be interpreted as the density of the compressed gas behind the shock, much higher than the density of the undisturbed interstellar medium (ISM), typically $\sim$1 cm$^{-3}$ in the spiral arms of a galaxy. 

From the dereddened H$\beta$ flux, at the distance of M\,51, we calculate an emitted luminosity $L_{{\rm{H}}\beta} \approx 3.7 \times 10^{36}$ erg s$^{-1}$. Knowing the characteristic temperature and density of the emitting gas, and using the H$\beta$ volume emissivity for Case B recombination \citep{1989agna.book.....O}, we infer an emitting volume $V_e \approx 8 \times 10^{56}$ cm$^3$. This is much lower than the volume inferred from the appearance of the nebula in the {\it HST} image, that is (roughly) an elongated bubble with a size of $\approx$30 $\times$ 100 pc corresponding to a volume of $\approx$10$^{60}$ cm$^{3}$. Such a discrepancy is expected for a shock-ionized bubble, because the line-emitting gas occupies only a very thin layer behind the shock. For the inferred density and volume of the shocked layer, the mass of currently line-emitting gas is $\sim$200 $M_{\odot}$. The total gas mass originally present in the $10^{60}$ cm$^3$ region that has now been swept up in the bubble must have been $\sim$10$^3$ $M_{\odot}$ for ISM densities $\sim$1 cm$^{-3}$. 
The cooling timescale of the shocked gas from the immediate, hot post-shock temperature ($T > 10^6$ K) to the warm temperature range in which most of the optical line emission is produced ($T \sim 10^4$ K) is $\sim$10$^4$ years, and that is followed by a short recombination timescale ($\approx$10$^3$ years) of the warm gas \citep{2003adu..book.....D}; on the other hand, typical ages of shock-ionized ULX bubbles are $\sim$10$^5$ yr. Thus, it is likely that much of the gas that passed through the shock has already cooled, even if the ULX is continually supplying radiative and kinetic power.

We then investigated whether the lines are broad or narrow. From the bright night-sky lines [O {\sc i}] $\lambda$5577 and Hg {\sc i} $\lambda$4358, we find an instrumental full width at half maximum FWHM$_{\rm ins,b} = 2.95$ \AA\ for the blue spectrum. In the red spectrum, strong night-sky lines are seen at $\lambda = 6329$ \AA, $\lambda = 6498$ \AA, $\lambda = 6863$ \AA: from the average width of those lines, we measure FWHM$_{\rm ins,r} =4.76$ \AA. The intrinsic line width FWHM$_{\rm int}$ is obtained from the observed width FWHM$_{\rm obs}$, following the relation ${\rm {FWHM}}_{\rm int} = ({\rm {FWHM}}_{\rm obs}^2 - {\rm {FWHM}}_{\rm ins}^2)^{1/2}$. Gaussian fits to the H$\beta$ and H$\alpha$ line profiles give observed FWHM$_{\rm obs}$ of $\approx$3.7 and 4.0 \AA, respectively, corresponding to intrinsic widths of $140\pm 20$ km s$^{-1}$ and $110 \pm 20$ km s$^{-1}$. Other low-ionization lines have similar intrinsic widths: $115 \pm 20$ km s$^{-1}$ for [O {\sc i}] $\lambda$6300, $115 \pm 20$ km s$^{-1}$ for [N {\sc ii}] $\lambda \lambda$6548,6583, and $130 \pm 20$ km s$^{-1}$ for [S {\sc ii}] $\lambda \lambda$6716,6731. Thus, all low-ionization lines are significantly resolved, and are indicative of a shock velocity $v_{\rm s} \approx {\rm FWHM}_{\rm int} \approx 110$--130 km s$^{-1}$. Further evidence of line broadening is obtained from the intrinsic half-width at zero-intensity (HWZI) for the Balmer lines; we measure HWZI(H$\alpha$) $\approx 280$ km s$^{-1}$, and HWZI(H$\beta$) $\approx 250$ km s$^{-1}$. On the other hand, somewhat unexpectedly, high-ionization lines are slightly narrower than low-ionization ones. Both [O {\sc iii}] $\lambda 4959$ and [O {\sc iii}] $\lambda 5007$ have intrinsic FWHMs $= 90\pm 5$ km s$^{-1}$ and HWZI $\approx 200$ km s$^{-1}$. He {\sc ii} $\lambda 4686$ is even narrower, with an intrinsic FWHM $= 60 \pm 20$ km s$^{-1}$ and HWZI $\approx 180$ km s$^{-1}$.

Finally, we examined the more characteristic line ratios. A detailed modelling of the spectrum is beyond the scope of this paper; however, here we highlight the presence and strength of the high ionization lines [Ne {\sc v}] $\lambda$3426, with $I(\lambda 3426)/I({\rm{H}}\beta) = 0.16 \pm 0.01$, and He {\sc ii} $\lambda$ 4686, with $I(\lambda 4686)/I({\rm{H}}\beta) = 0.22 \pm 0.01$. Neither of those lines can be produced at such intensity by a relatively slow ($\approx$120 km s$^{-1}$) shock plus precursor. The combined presence of [Ne {\sc v}] $\lambda$3426 and He {\sc ii} $\lambda$ 4686 is the most intriguing feature of some ultracompact dwarf galaxies \citep{2012MNRAS.427.1229I}; the origin of the lines in those galaxies was unclear, but the most likely candidates were suggested to be fast radiative shocks or an AGN contribution. In M\,51 ULX-1, we have the rare opportunity to identify a source of such line emission. Moreover, the emission from [Ne {\sc iii}] $\lambda$3869, [O {\sc iii}] $\lambda$4959,5007, and [N {\sc ii}] $\lambda$6548,6583 is remarkably strong, with $I(\lambda 3869)/I({\rm{H}}\beta) = 0.92 \pm 0.05$, $I(\lambda 4959)/I({\rm{H}}\beta) = 2.1 \pm 0.1$ and $I(\lambda 6548)/I({\rm{H}}\beta) = 1.27 \pm 0.05$. If entirely due to shock-ionization, those line ratios are approximately consistent only with shock velocities $\sim$500--600 km s$^{-1}$. Based on the shock plus precursor models of \citet{2008ApJS..178...20A}, at this range of velocity, these line ratios are only weakly dependent on the unperturbed ISM density, metal abundance and magnetic field. We do not see such high-velocity wings in the LBT spectra; howewer, this could be due to our viewing angle of the collimated outflow in the plane of the sky, with the direction of the fast outflow almost perpendicular to our line of sight.

\begin{figure}
\centering
\includegraphics[width=0.48\textwidth]{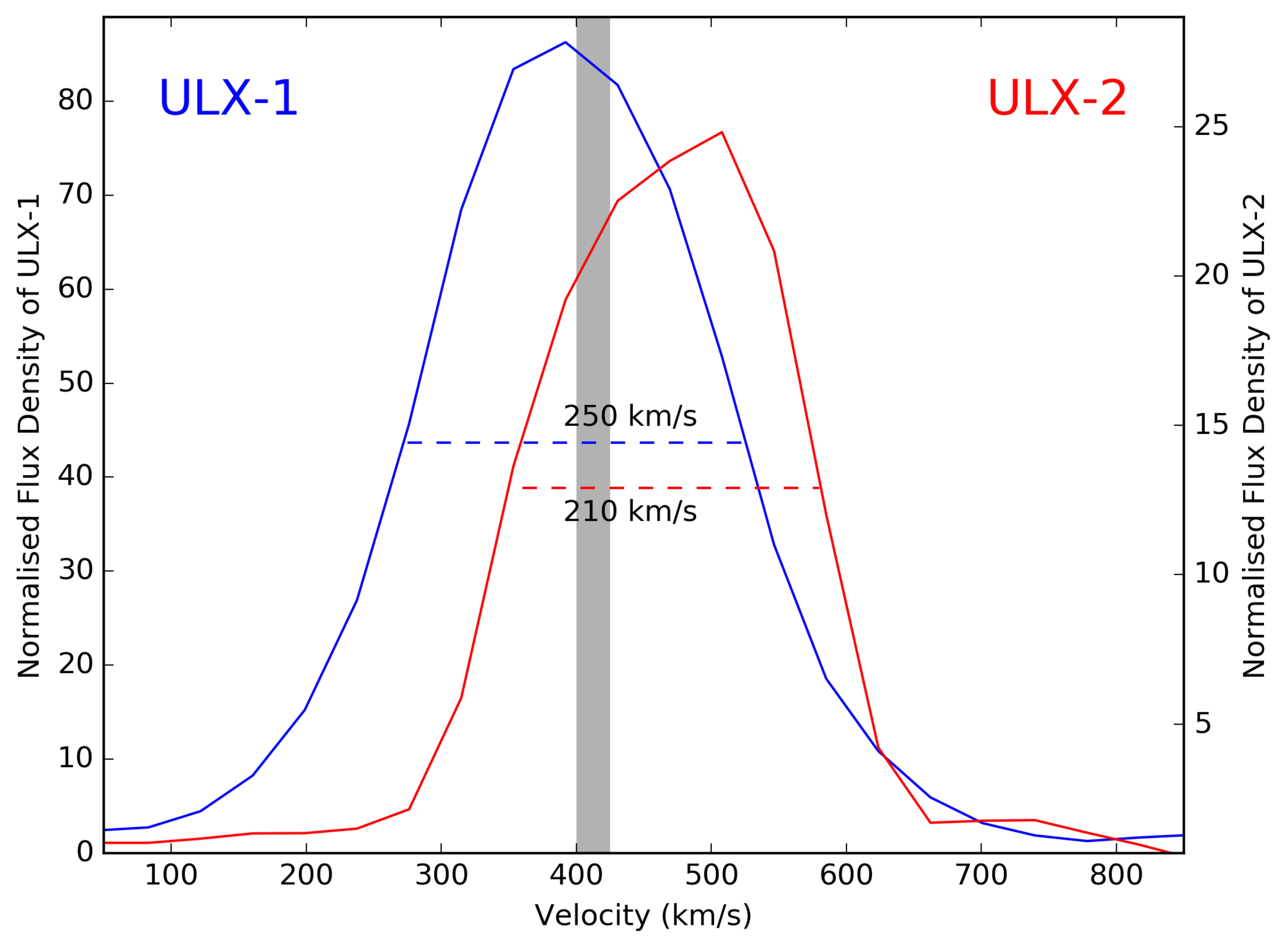}
\includegraphics[width=0.48\textwidth]{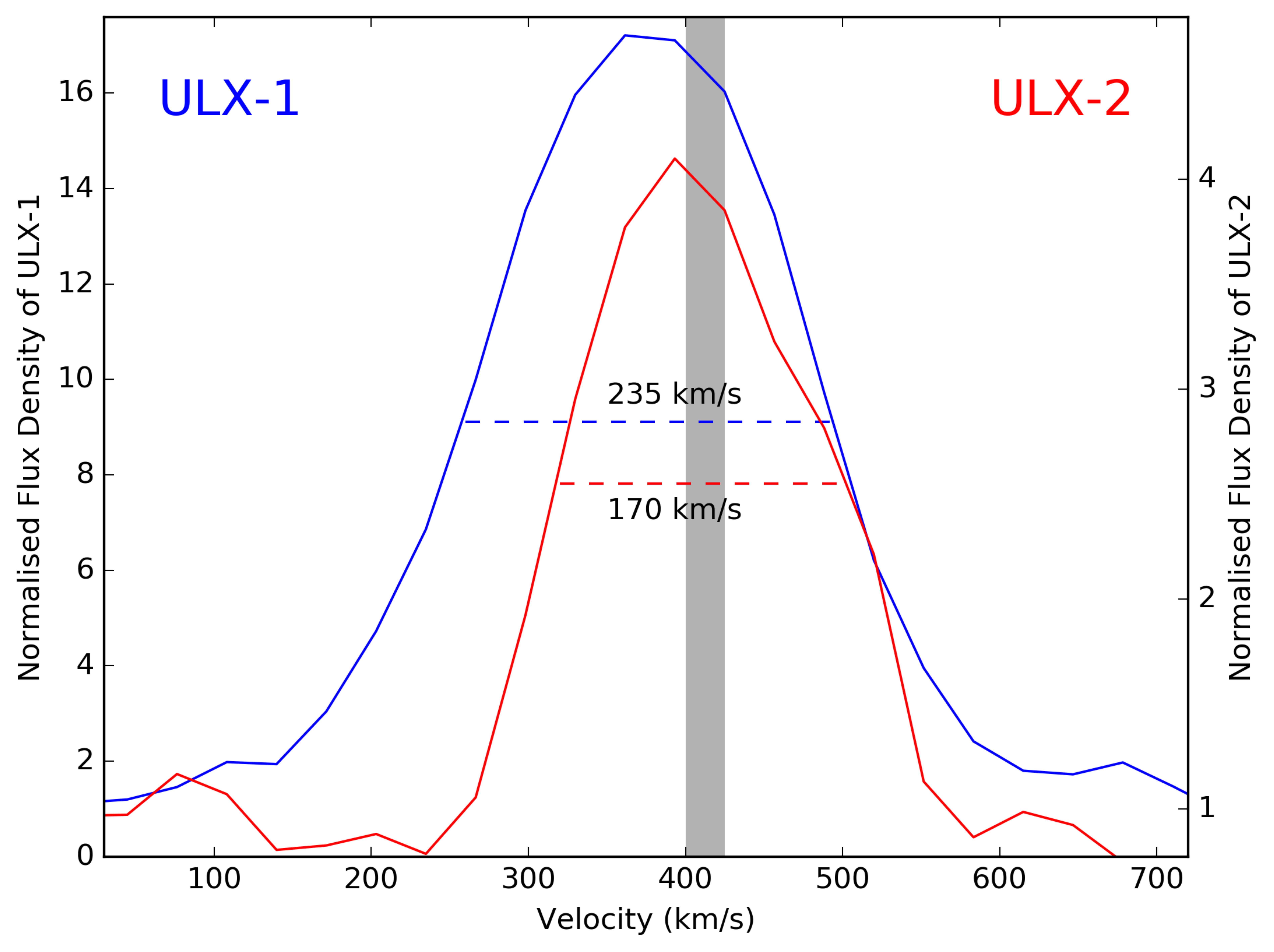}
\includegraphics[width=0.48\textwidth]{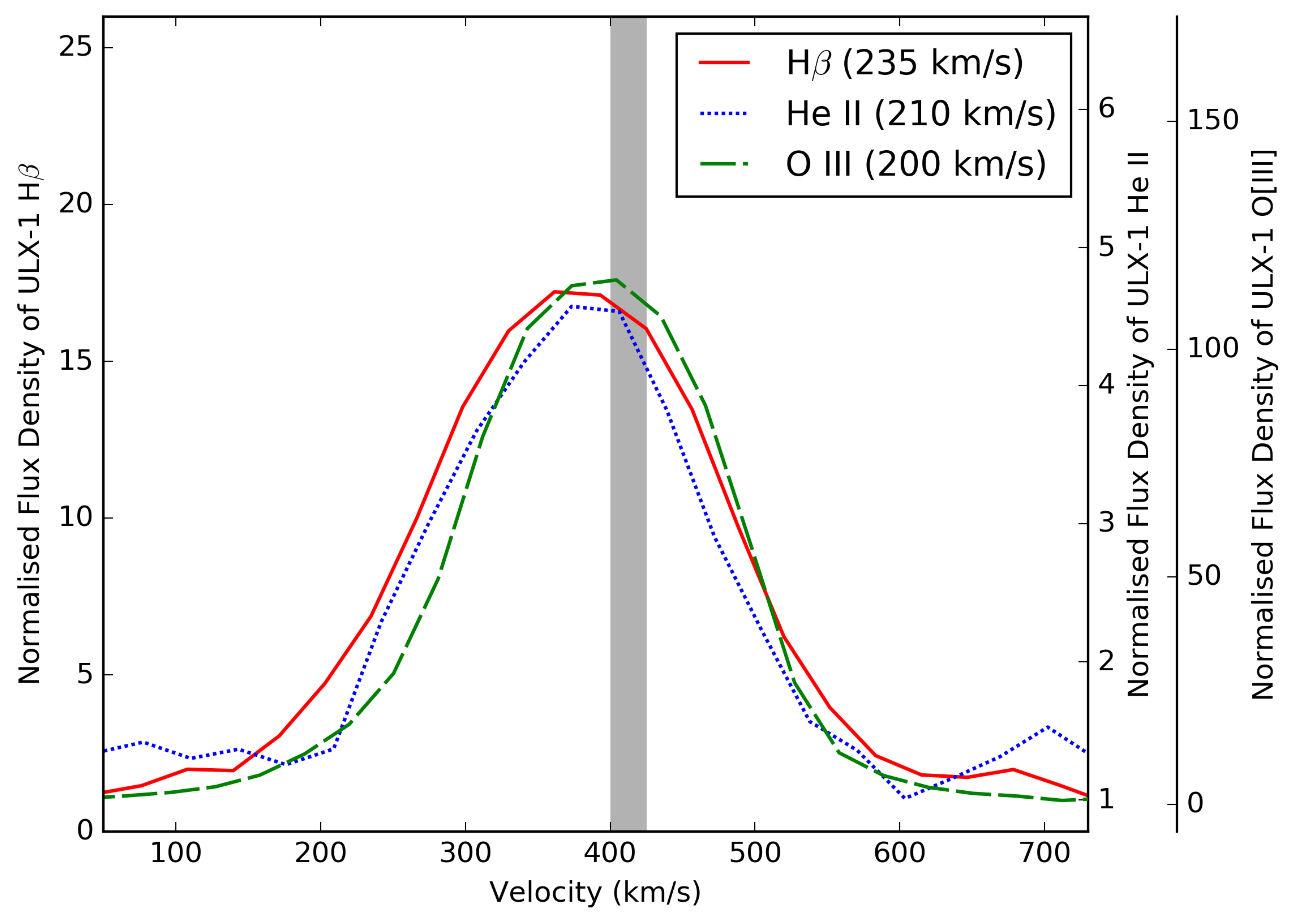}
\vspace{-0.4cm}
 \caption{Top panel: H$\alpha$ line profile for the nebulae around ULX-1 (blue) and around ULX-2 (red). The shaded band indicates the systemic velocity expected in the M\,51 spiral arm at the location of ULX-1 and ULX-2 ($\approx$400--425 km s$^{-1}$, from \citet{2013MNRAS.433.1837V} and \citet{2007ApJ...665.1138S}). The FWHM of the two lines is also overplotted; both the FWHM and the FWZI are higher in the ULX-1 nebula. Middle panel: same as the top panel, but for the H$\beta$ line; again, the line from the ULX-1 nebula is broader. Bottom panel: comparison between the line profiles of He {\sc ii} $\lambda 4686$, [O {\sc iii}] $\lambda 5007$, and H$\beta$ (plotted with the same peak intensity for convenience) for the ULX-1 nebula: the relatively small FWHM and lack of broad wings in the higher-ionization lines are inconsistent with the high shock velocity that would be required to produce such lines entirely from shocks, and suggests a significant contribution to those lines from X-ray photo-ionization.}
  \label{vel_lines_im}
  \vspace{0.3cm}
\end{figure}

\subsubsection{ULX-2 nebula}

The spectrum of the nebula around ULX-2 (Table 1 and Figure \ref{ulx2_opt_lines_im}) is clearly different from that of the ULX-1 nebula. From an analysis of the line width, we find that all lines are consistent with the instrumental widths (see Section 3.2.1). That is evidence that the gas is photo-ionized rather than shock-ionized. The [N {\sc ii}] and [S {\sc ii}] lines are much weaker, relative to H$\alpha$, with flux ratios ([N {\sc ii}] $\lambda$6548 + [N {\sc ii}] $\lambda$6583)/H$\alpha$ $\approx$ 0.47 and ([S {\sc ii}] $\lambda$6716 + [S {\sc ii}] $\lambda$6731)/H$\alpha$ $\approx$ 0.28. This is again consistent with a photo-ionized H {\sc ii} region. 

Again we tried using standard diagnostic line ratios for temperature and density. Unfortunately, the [O {\sc iii}] $\lambda$4363 line is too faint to provide meaningful temperature constraints. Instead, for the [S {\sc ii}] line doublet, we find an intensity ratio [S {\sc ii}] $I(\lambda 6716)/I(\lambda 6731) \approx 1.45 \pm 0.05$. For temperatures $\sim$10$^4$ K, this ratio corresponds to the low-density limit, $n_e < 10$ cm$^{-3}$, consistent with the standard ISM density of $\sim$1 cm$^{-3}$. Given the close proximity and similar environment of ULX-1 and ULX-2, it is plausible to assume that was also the unperturbed ISM density around ULX-1, before the gas was compressed by the shock front. 

With $\log$ ([N {\sc ii}] $\lambda$6583/H$\alpha$) $\approx -0.51$, $\log$ ([S {\sc ii}] $\lambda$6716 $+$ $\lambda$6731/H$\alpha$) $\approx -0.55$, and $\log$ ([O {\sc iii}] $\lambda$5007/H$\alpha$) $\approx -0.02$, the nebula sits along the sequence of H {\sc ii} regions photo-ionized by stellar emission \citep{2002ApJS..142...35K, 1987ApJS...63..295V,1981PASP...93....5B}. The lack of high-ionization lines such as He {\sc ii} $\lambda$4686 and [Ne {\sc v}] $\lambda$3426 (detected in the ULX-1 nebula, Section 3.2.1) is further evidence that most of the ionizing photons come from the stellar population rather than the ULX. Non-detection of [O {\sc i}] $\lambda$6300 suggests the lack of an extended low-ionization zone, which we would expect to find if the nebula were ionized by soft X-ray photons from ULX-2. 

Using nitrogen and oxygen lines as metallicity indicators, we run into the well-known discrepancy \citep{2003ApJ...591..801K,2004ApJ...615..228B} between the ``direct" metallicity calibration based on oxygen temperatures, and the ``indirect" calibration based on photo-ionization models. The direct $T_e$-based measurements require detection of faint auroral lines ({\it e.g.}, [O {\sc iii}] $\lambda$4363; \citealt{2003ApJ...591..801K}), while photo-ionization models constrain the metal abundance from the intensity of stronger lines, for example the ratio between [N {\sc ii}] $\lambda$6584 and [O {\sc ii}] $\lambda \lambda$3726,3729 \citep{2002ApJS..142...35K}. It was noted \citep{2004ApJ...615..228B} that indirect photo-ionization abundances of H {\sc ii} regions in nearby spiral galaxies are systematically higher than $T_e$-measured abundances by a factor of 2--3. For our present study, we cannot directly measure $T_e$ abundances for the ULX-2 nebula; we have to rely on stronger lines (Table 2) and use either the direct or the indirect calibration between those line ratios and the metal abundance. Applying the $T_e$-based calibration, from our measured values of $\log$ ([N {\sc ii}] $\lambda$6583/H$\alpha$) and of $\log$ [([O {\sc iii}] $\lambda$5007/H$\beta$)/([N {\sc ii}] $\lambda$6583/H$\alpha$)] we find $\log$ (O/H) $+ 12 \approx 8.6 \pm 0.1$, that is about 20\% below solar (using the definition of solar abundance from \citealt{2001ApJ...556L..63A}). This is the same as the metal abundance directly measured for several other H {\sc ii} regions in the same spiral arm of M\,51, at similar radial distances from the galactic nucleus \citep{2015ApJ...808...42C,2004ApJ...615..228B}. On the other hand, based on the indirect calibration of \cite{2002ApJS..142...35K}, \cite{2004ApJ...617..240K}, and \cite{2005ApJ...631..231P}, the same line ratios correspond to $\log$ (O/H) $+ 12 \approx 9.1 \pm 0.1$, that is $Z \approx 2.5 Z_{\odot}$. A similar discrepancy also occurs if we use the so-called $R_{23}$ metallicity indicator, that is the intensity ratio between ([O {\sc ii}] $\lambda \lambda$ 3726,3729 $+$ [O {\sc iii}] $\lambda \lambda$4959,5007) over H$\beta$ \citep{1979MNRAS.189...95P, 1984MNRAS.211..507E, 2002ApJS..142...35K, 2008ApJ...681.1183K}. The traditional calibration of this ratio based on photo-ionization models gives $\log$ (O/H) $+ 12 \approx 9.1$, 3 times higher than the value  derived from the direct measurements of \cite{2003ApJ...591..801K} and \cite{2004ApJ...615..228B} (see in particular Fig.~7 in the latter paper). Previous work in the literature supporting a super-solar metallicity for the M\,51 disk \citep{2010ApJS..190..233M, 1994ApJ...420...87Z} was also calibrated on photo-ionization models. Resolving the systematic discrepancy between the two metallicity calibrations will be important for population synthesis models of ULXs, because the ratio between neutron stars and black holes produced from stellar collapses increases strongly at super-solar metallicities \citep{2003ApJ...591..288H}.


\subsection{Radio counterpart of ULX-1}

\begin{figure}
\centering
\includegraphics[width=0.48\textwidth]{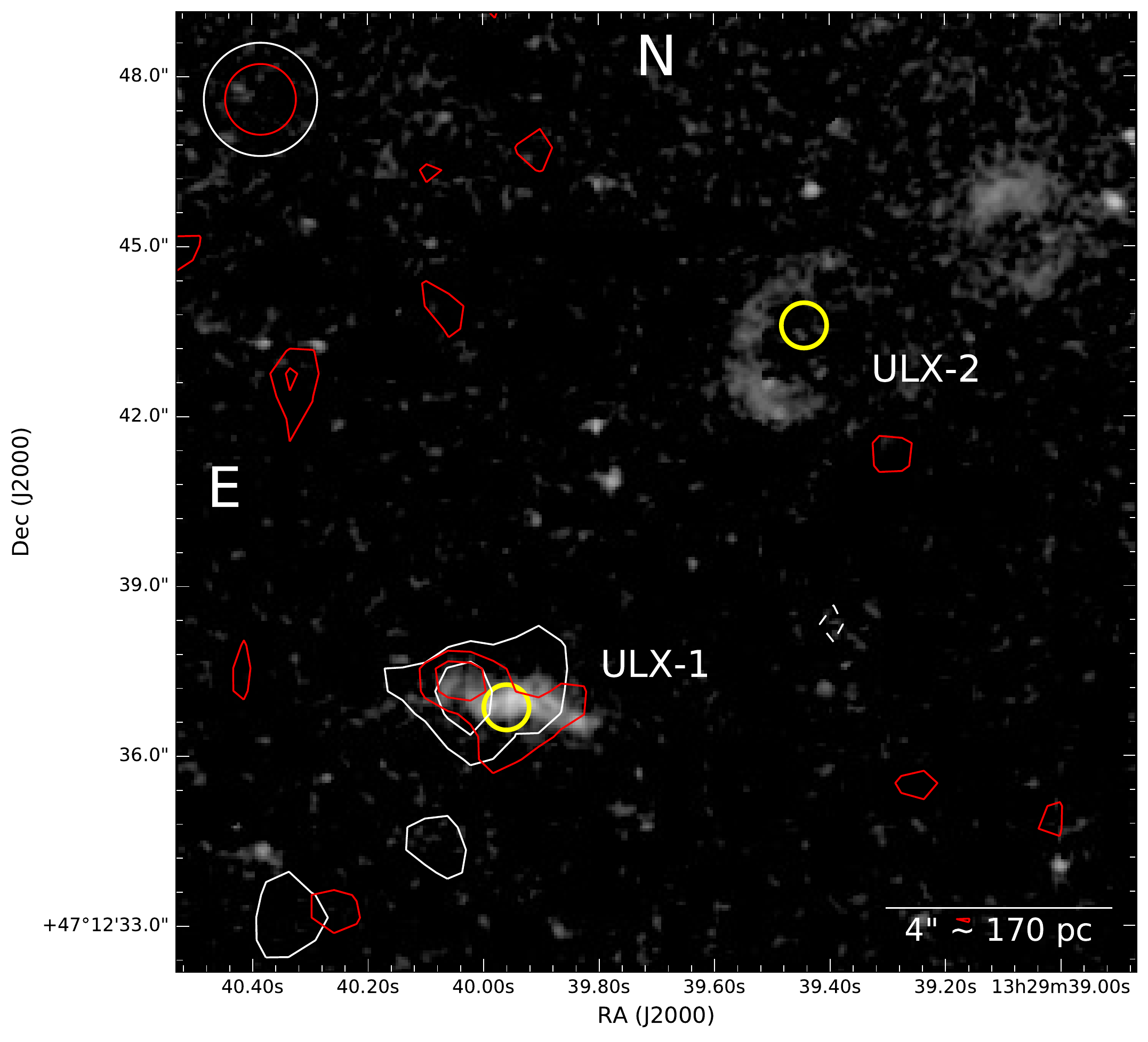}
\includegraphics[width=0.48\textwidth]{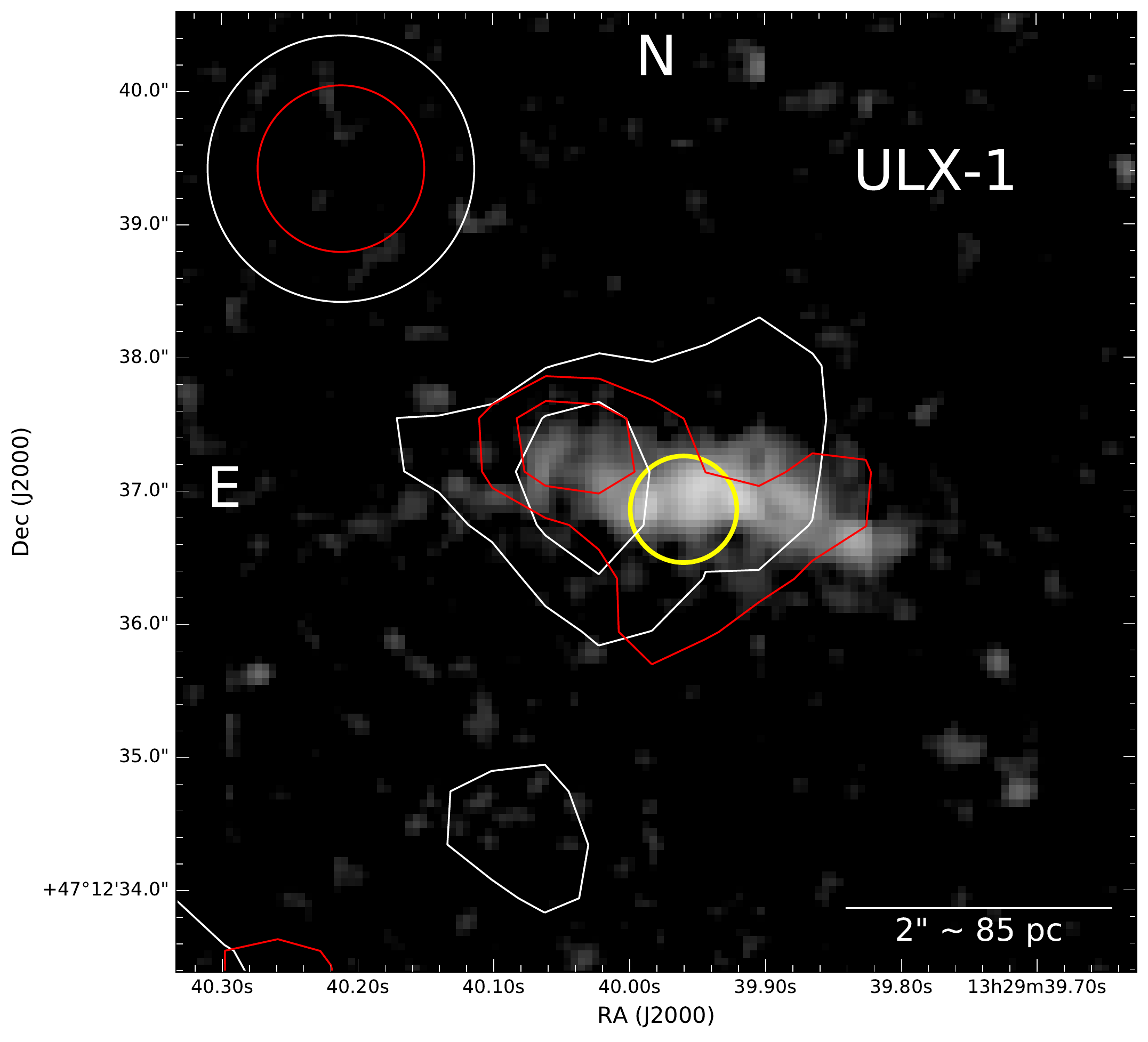}
 \caption{Top panel: {\it HST}/ACS continuum-subtracted F658N image of the ULX field, with VLA radio contours overplotted. White and red contours represent the 4.9-GHz and 1.4-GHz flux densities, respectively. Both sets of contours have levels of $-2\sqrt{2}\sigma$, $2\sqrt{2}\sigma$ and $4\sigma$, where $\sigma$ is the respective noise level ($\sigma_{4.9} = 7.3$ $\mu$Jy\perbeam, $\sigma_{1.4} = 9.2$ $\mu$Jy\perbeam). Beams for both frequencies are indicated in the top left corner of the image. The yellow circles represent the {\it Chandra} positions of ULX-1 and ULX-2 with 0$^{\prime\prime}$.4 uncertainty. Bottom panel: zoomed-in view of the ULX-1 bubble.}
  \label{radio_im}
  \vspace{0.3cm}
\end{figure}

The VLA+Effelsberg data show a faint radio source near ULX-1, detected both at 1.4 GHz and at 4.9 GHz, however not detected at 8.4 GHz. In the 1.4 and 4.9 GHz radio images the source is marginally resolved, elongated in the east-west direction (Figure \ref{radio_im}), that is the same long axis of the optical nebula. The position of the peak intensity in the radio source is slightly offset, by $\approx$1$^{\prime\prime}$, to the east of the {\it Chandra} position and of the central region of the ULX bubble in the {\it HST} image. Such an offset is significantly larger than the astrometric uncertainty (Section 2.4) of the X-ray and optical images. Thus, we speculate that the peak radio emission comes from a hot spot where the eastern jet interacts with the ISM, rather than being associated with the core. From the primary-beam corrected data, we determined peak brightnesses $f_{1.4\,\GHz}=(44.8\pm9.2)\,\micro\jansky$\perbeam, $f_{4.9\,\GHz}=(29.8\pm7.3)\,\micro\jansky$\perbeam, and an 8.4-GHz upper-limit of $f_{8.4\,\GHz}<54\,\micro\jansky$\perbeam. The integrated fluxes are  $f_{1.4\,\GHz}=(110\pm23)\,\micro\jansky$ and $f_{4.9\,\GHz}=(52.7\pm12.4)\,\micro\jansky$. At the distance of M\,51, this corresponds to a 4.9-GHz luminosity\footnote{The luminosity is defined as $L_{\nu} = 4\pi d^2\nu f_{\nu}$.} $L_{4.9\,\GHz} = (2.3 \pm 0.5) \times 10^{34}$ erg s$^{-1}$. From the integrated fluxes, we determine a spectral index\footnote{Defined as $f_{\nu} \propto \nu^{\alpha}$} $\alpha=-0.6\pm0.3$, consistent with the optically-thin synchrotron emission expected from hot spots. Radio flux may also come from free-free emission associated with the ionized bubble: however, based on the Balmer emission, we only expect a free-free flux density of $\approx$1.5 $\micro\jansky$ at 4.9 GHz (\citealt{1986A&A...155..297C}, Appendix A), entirely negligible. The radio emission may also be a result of a supernova remnant; however, the highly elongated morphology of the optical nebula suggests that we are looking at a strongly collimated outflow, thus more likely driven by the ULX.

We do not detect any radio emission associated with ULX-2 or its surrounding nebula. We can only place 3$\sigma$ upper limits on the peak flux densities: $f_{1.4\,\GHz}<28\,\micro\jansky$\perbeam, $f_{4.9\,\GHz}<22\,\micro\jansky$\perbeam, and $f_{8.4\,\GHz}<54\,\micro\jansky$\perbeam. This corresponds to an upper limit to the 4.9-GHz luminosity $L_{4.9\,\GHz} < 10^{34}$ erg s$^{-1}$. The expected free-free emission from the H {\sc ii} region around ULX-2 is $\approx$2 $\micro\jansky$ at 4.9 GHz \citep{1986A&A...155..297C}, well below the detection limit.

\section{Discussion}

Although the X-ray luminosity of ULX-1 and ULX-2 is only moderately above $10^{39}$ erg s$^{-1}$, the two sources have interesting properties that can help us understand the behaviour of super-critical accretion. Both sources are eclipsing \citep{2016ApJ...831...56U}, showing sharp transitions in which the companion star occults the compact object. Eclipsing sources are very rare among ULXs and X-ray binaries in general. The presence of periodic eclipses enables a number of binary parameters to be constrained ({\it e.g.} binary period, inclination angle, mass function, size of the X-ray emitting region). This makes these objects prime targets for follow-up multiwavelength observations. Here we have shown that both ULXs are associated with ionized gas nebulae, with different optical spectral properties. We have also shown that ULX-1 (but not ULX-2) has a radio counterpart (usually taken as a signature of jet activity). We shall now discuss the physical interpretation of the sources, putting together the various clues from these multiband observations.

\subsection{Energetics of the two nebulae}

As we have shown in Section 3.2, the most obvious difference between the two nebulae is that one is dominated by shock ionization (with possible additional contribution from X-ray photo-ionization), the other by near-UV photo-ionization. We shall discuss here how much kinetic and radiative energy is required to produce the luminosity observed from the two nebulae, and whether the two ULXs are the origin of such power. 

\subsubsection{Mechanical power of ULX-1}

From standard bubble theory \citep{1977ApJ...218..377W}, the mechanical power $P_{\rm jet}$ inflating the bubble is equal to $\approx$77/27 of the total radiative luminosity $L_{\rm rad}$. The luminosity in the H$\beta$ line ($L_{\rm H\beta}$) is between $\approx$4 $\times 10^{-3}$ and $\approx$7 $\times 10^{-3}$ times the total radiative luminosity, for all shock velocities $100 \la v_{\rm s} \la 500$ km s$^{-1}$ (from the set of pre-run shock-ionization {\sc mappings} III models for equipartition magnetic field and either solar or twice-solar abundance; \citealt{2008ApJS..178...20A}). Thus, $P_{\rm jet} \approx (400$--$650) \times L_{\rm {H}\beta}$. Knowing that $L_{\rm {H}\beta} \approx 3.7 \times 10^{36}$ erg s$^{-1}$, we infer a mechanical power $P_{\rm {jet}} \approx (1.5$--$2.5) \times 10^{39}$ erg s$^{-1}$. This value is comparable to the X-ray luminosity of ULX-1 \citep{2016ApJ...831...56U}, taking into account the uncertainty in the viewing angle and therefore the geometric projection factor for the disk emission. It is also similar to the jet power inferred for SS\,433 \citep{2017A&A...599A..77P, 2004ASPRv..12....1F}. 
There is no direct conversion between jet power and luminosity of the radio counterpart because the latter also depends on other factors such as the ISM density and magnetic field, and the fraction of kinetic power carried by protons and ions; however, it is interesting to note that the 5-GHz luminosity $L_{4.9\,\GHz} \approx 2.3 \times 10^{34}$ erg s$^{-1}$ is within a factor of two of the radio luminosity of other jetted ULXs such as Holmberg II X-1 \citep{2014MNRAS.439L...1C, 2015MNRAS.452...24C} and NGC\,5408 X-1 \citep{2012ApJ...749...17C, 2007ApJ...666...79L, 2006MNRAS.368.1527S}. Thus, the radio luminosity of the M\,51 ULX-1 nebula seems to correspond to ``average" jetted ULX properties: it is a factor of 5 more radio luminous than SS\,433, but an order of magnitude fainter than the most exceptional radio bubbles in the local universe, namely NGC\,7793 S26 \citep{2010Natur.466..209P, 2010MNRAS.409..541S} and IC\,342 X-1 \citep{2012ApJ...749...17C}, which are also a few times larger. The radio spectral index $\alpha=-0.6 \pm 0.3$ confirms that the radio counterpart is dominated by optically-thin synchrotron emission, as expected for jet lobes.

An alternative method to estimate the jet power \citep{1977ApJ...218..377W} is to assume that the moderate shock velocity $\approx$100--150 km s$^{-1}$ inferred from the FWHM of the Balmer lines also corresponds to the advance speed of the jet-powered forward shock into the ISM. In that scenario, the characteristic age of the bubble is $\approx$3 $\times 10^5 v^{-1}_{100}$ yr and the jet power $P_{\rm {jet}} \approx 1.0 \times 10^{39} v^{3}_{100} n_e$ erg s$^{-1}$, where $v_{100} \equiv v_{\rm s}/(100{\rm{~km~s}}^{-1})$ and $n_e$ is the ISM number density. This is consistent with the other jet power estimate.

There is clearly a strong discrepancy between the moderately low FWHM of all the lines ($\sim$100 km s$^{-1}$) and the high shock velocity ($\sim$500 km s$^{-1}$) that would be needed to produce the higher ionization lines with the observed ratios; such high shock velocities are also inconsistent with the observed HWZIs and the moderate gas temperature inferred from the [S {\sc ii}] lines. Even if we account for the fact that about half of the high-ionization line emission occurs in the precursor region (low turbulent velocity), the discrepancy remains difficult to explain. One possibility is that there is a contribution from X-ray photo-ionization, from ULX-1. The source appears almost edge-on to us, but if, as expected, its polar funnel is aligned with the major axis of the bubble ({\it i.e.} along the jet), there is stronger direct X-ray irradiation and lower intrinsic absorption along that direction. The fact that high-ionization lines are slightly narrower than low-ionization lines supports the idea that the former are partly enhanced by X-ray photo-ionization. Another possibility is that the shocked gas does have a faster expansion velocity (comparable with the shock velocity) along the direction of the jet (which is located approximately in the plane of the sky for us) and a slower expansion in the other two directions perpendicular to the jet axis; in that case, the line width observed by us only reflects the lateral expansion speed. Soft thermal X-ray emission is a possible test for the presence of fast shocks: a shock velocity $\approx$500 km s$^{-1}$ corresponds to a shock temperature $\approx$0.3 keV. Soft thermal emission was indeed detected in the X-ray spectrum of ULX-1 \citep{2016ApJ...831...56U}; unfortunately, the location of the source away from the aimpoint in the {\it Chandra} observations, and the low spatial resolution of {\it XMM-Newton}, do not enable us to determine whether the thermal-plasma component comes only from the central point-like source or from the larger bubble. In the former case, the X-ray line emission would be directly linked to the disk outflow around the ULX; in the latter case, it would be from the jet/ISM interaction. New {\it Chandra} observations with the source at the aimpoint could answer this question.

\subsubsection{Radiative power of ULX-2} \label{ulx2_sec}

The optical spectrum of the nebula around ULX-2 is consistent with a typical H {\sc ii} region; this would mean that it is only a coincidence that a ULX is apparently located inside the nebula, without contributing to its ionization. To determine whether this is the case, we start by measuring its extinction-corrected H$\beta$ luminosity: $L_{\rm {H}\beta} \approx 4.0 \times 10^{36}$ erg s$^{-1}$. At $T_e \approx 10,000$ K, the emission of one H$\beta$ photon requires $\approx$8.5 ionizing photons above 12.6 eV \citep{1989agna.book.....O}, and below $\approx$0.2 keV (above which energy the ionization cross section becomes too low). Therefore, the observed nebular luminosity implies an ionizing photon flux $Q({\rm H}^0) \approx 8 \times 10^{48}$ photons s$^{-1}$. From the best-fitting parameters in the solar-metallicity stellar tracks, we estimate that the two blue stars in the error circle of ULX-2 have effective temperatures of $\approx$20,000 K and $\approx$17,000 K, and radii of $\approx$25 $R_{\odot}$ and $\approx$36 $R_{\odot}$, respectively. Approximating the spectra of the two stars as blackbodies, we expect a combined ionizing flux of $\approx$1.4 $\times 10^{48}$ photons s$^{-1}$ above 12.6 eV. Adding a few other blue stars inside the outer boundary of the nebula still leaves us a factor of 3 short of the required ionizing flux. However, using only the bright blue supergiants for this type of photon accounting can be misleading. For example, the addition of a single main-sequence mid-type O star (O6V or O7V), with a temperature $\approx$35,000--37,000 K, would suffice to provide the remaining ionizing flux \citep{2008MNRAS.389.1009S}. Such O-type dwarfs have an absolute brightness $M_V \approx -4.5$ mag, negligible compared with the characteristic absolute brightness $M_V \approx -6$ mag of the B4--B6 supergiants around ULX-2. Thus, if one of the four supergiants visible around ULX-2 (Figure \ref{field_im}, bottom right) is in a binary system with a mid-type O star, the supergiants will determine the optical colours and $V$-band luminosity, but the O star will provide most of the ionizing UV flux. Another way to tackle the problem is to use {\sc starburst99} \citep{1999ApJS..123....3L,2014ApJS..212...14L} simulations: we find that for a single population age of 7 Myr, a stellar mass of $\approx$10$^4 M_{\odot}$ would be required to provide the necessary ionizing flux. 
Adding a dust reddening correction for the M\,51 halo (as discussed in Sections 2.2 and 3.1) would increase the intrinsic ionizing luminosity of the stars by $\approx$50\%, making it easier to meet the energy requirements for the H {\sc ii} region.

Estimating whether there may also be a direct photo-ionizing contribution from ULX-2 is not trivial. The X-ray source is seen at a high inclination angle $\theta$, which implies that an isotropic conversion of observed high-energy flux into emitted luminosity may underestimate the disk emission; a conversion factor $\approx$2$\pi \, d^2/\cos \theta$ is more appropriate than a factor of $4\pi \, d^2$. Taking this into account, we can always find a sufficiently high viewing angle (in particular, $\theta \ga 80^{\circ}$) to increase the true emitted disk luminosity of the source. To boost the number of UV photons, we must also assume that the X-ray source is surrounded by a large disk that intercepts and reprocesses more than 1\% of the X-ray flux. Thus, an ionizing flux $Q({\rm H}^0) \approx 8 \times 10^{48}$ photons s$^{-1}$ is consistent with the X-ray and optical luminosity of ULX-2. However, the problem of this irradiated-disk scenario is that it also produces copious amount of photons at energies $>54$ eV, which leads to the emission of He {\sc ii} $\lambda 4686$ via recombination of He$^{++}$ into He$^{+}$. No He {\sc ii} $\lambda 4686$ emission is detected in the LBT spectra of ths nebula, with an upper limit on its flux of $\approx$10$^{-16}$ erg cm$^{-2}$ s$^{-1}$, corresponding to an emitted photon flux $\la$2 $\times 10^{47} \lambda 4686$ photons s$^{-1}$. At $T_e \sim 10,000$ K, it takes $\approx$4.2 primary ionizing photons $>$54 eV to produce one $\lambda 4686$ photon \citep{1986Natur.322..511P, 1989agna.book.....O}: thus, the flux of $E > 54$ eV photons seen by the nebular gas must be $\la 8 \times 10^{47}$ photons s$^{-1}$ $\approx 0.1 Q({\rm H}^0)$.  Typical models of irradiated disks ({\it e.g., diskir} in {\sc xspec}) luminous enough to produce the required number of photons $>$13.6 eV will also predict too many photons $>$54 eV (typically, the latter are predicted to be $\approx$1/4 of the former, rather than $\la$1/10, as required by the observed spectrum). Given this evidence, we conclude that the ionized nebula projected around ULX-2 is a typical H\,{\sc ii} region and the ULX, whether located inside the nebula or not, is not its major ionizing source.

\subsection{Radio brighter and radio fainter ULX bubbles} \label{sec_ulx_dis}

ULX bubbles are a crucial tool to estimate the emitted power (radiative and kinetic luminosity) of the compact object, averaged over the characteristic cooling timescale of the gas. Moreover, they enable us to identify compact objects with likely super-critical accretion that currently appear X-ray faint, either because they are in a low state, or because they are collimated away from us, or because the direct X-ray emission along our line of sight is blocked by optically thick material. As more ULX bubbles get discovered or recognized, we are starting to see that there is no simple correlation between the X-ray luminosity of the central source and the radio or Balmer line luminosity of the bubble. The differences between M\,51 ULX-1 and ULX-2 illustrated in this paper are a clear example. This is also the case when we compare the M\,51 ULX-1 bubble with ULX bubbles in other galaxies. For example, the collisionally ionized ULX bubble around Holmberg IX X-1 \citep{2011ApJ...734...23G,2008RMxAA..44..301A,2008AIPC.1010..303P,2002astro.ph..2488P} has an H$\alpha$ luminosity $\approx$10$^{38}$ erg s$^{-1}$ \citep{2008RMxAA..44..301A,1995ApJ...446L..75M,1994ApJ...427..656M}, an order of magnitude higher than that of the M\,51 ULX-1 bubble, consistent with a mechanical power of $\approx$10$^{40}$ erg s$^{-1}$; the central ULX powering the Holmberg IX bubble has an X-ray luminosity $L_{\rm{X}} \approx 2 \times 10^{40}$ erg s$^{-1}$ (\citealt{2017ApJ...839..105W} and references therein), again an order of magnitude higher than M\,51 ULX-1. However, the Holmberg IX ULX bubble has a 1.4-GHz luminosity of $\approx$2 $\times 10^{34}$ erg s$^{-1}$ (our estimate based on the radio maps of \citealt{1989A&A...217...17K}), similar to the radio luminosity of M\,51 ULX-1. Conversely, there are ULXs with more luminous radio nebulae (such as Holmberg II X-1 and NGC\,5408 X-1; see Section 4.1.1 for references) and proof of powerful jets, but with relatively weak or no collisionally ionized gas detectable from optical lines. The different ratio of synchrotron radio luminosity over mechanical power (or radio luminosity over total accretion power) in different ULXs may be due to various factors: i) a different initial kinetic energy distribution between a collimated, relativistic jet and a slower, more massive disk wind; ii) different composition of the jet (leptonic or baryonic); iii) different amount of entrainment of ISM matter along the jet path (related to different ISM densities). A general classification of ULX bubbles in terms of their multiband luminosity ratios is beyond the scope of this work.

For the ULXs that do show radio evidence of extended jets, typical 5-GHz luminosities span an order of magnitude between $\approx$2$\times 10^{34}$ erg s$^{-1}$ in M\,51 ULX-1 to $\approx$2$\times 10^{35}$ erg s$^{-1}$ in NGC\,7793-S26 \citep{2010MNRAS.409..541S} and IC\,342 X-1 \citep{2012ApJ...749...17C}; the characteristic age of the ULX bubbles is $\sim$ a few $10^5$ yr. This luminosity range corresponds to a range of observed fluxes $f_{\rm{5GHz}} \approx (30$--$300)\,d^{-2}_{10} \mu$Jy, where $d_{10}$ is the source distance in units of 10 Mpc. The synchrotron radio flux observed from microquasar lobes during the active expansion phase, that is when the jet is still inflating them, was modelled by \citet{2002A&A...388L..40H} as $f_{\rm{5GHz}} \sim 10 \,n_e^{0.45} \, P_{\rm jet,39}^{1.3} \, \left[2\varphi^{3/4}/(1+\varphi)\right] \, t_5^{0.4}\,d^{-2}_{10} \, \milli$Jy, where $\varphi$ is the ratio between the magnetic pressure and the gas pressure at the base of the jet, $P_{\rm jet,39}$ is the jet power in units of $10^{39}$ erg s$^{-1}$, $t_5$ is the age of the jet active phase in units of $10^5$ yr, and $n_e$ is the number density in the ISM. 
While the \citet{2002A&A...388L..40H} scaling relation works well for the radio nebula around the neutron star microquasar Cir X-1, it overestimates the synchrotron radio luminosity of the bubble around the BH microquasar Cyg X-1 by at least two orders of magnitude \citet{2005Natur.436..819G}; it also over-estimates the synchrotron radio luminosity of ULX bubbles (given the input mechanical power inferred from line-emission studies) by a similar amount. For Cyg X-1, this may be the result of a significant fraction of the jet energy being stored in baryons that do not radiate \citep{2005Natur.436..819G,2006ApJ...636..316H}. A similar explanation may apply to jet-driven ULX bubbles \citep{2010MNRAS.409..541S}.

\section{Conclusions}

We presented the discovery and multiband study of two ionized nebulae spatially associated with the two eclipsing ULXs in M\,51, using new and archival data from {\it Chandra}, the LBT, the HST, and the VLA. The nebula around ULX-1 has a very elongated morphology, indicative of a jet-driven bubble. We showed that the optical emission lines are collisionally ionized, and that there is a spatially resolved radio counterpart. Such findings are consistent with the jet interpretation. We know from X-ray photometry that ULX-1 is seen almost edge on; therefore the jet is likely propagating in the plane of the sky, which explains the lack of systemic velocity shifts observed between the eastern and western ends of the optical bubble. The mechanical power of the jet estimated from the H$\beta$ luminosity, $P_{\rm jet} \approx 2 \times 10^{39}$ erg s$^{-1}$, agrees with the power estimated from the FWHM of the lines (proportional to the shock velocity) and the characteristic age of the source ($t \sim 10^5$ yr). However, high-ionization lines (specifically, [Ne {\sc v}] $\lambda 3426$, He {\sc ii} $\lambda 4686$, and a few [O {\sc iii}] lines) are too strong for the shock velocity $v_{\rm s} \sim 100$--150 km s$^{-1}$ inferred from the width of the low-ionization lines. Either there is also a shock component with $v_s \sim 500$ km s$^{-1}$, or the high-ionization lines are boosted by X-ray photo-ionization. We favour the latter scenario because the high-ionization lines are slightly narrower than the Balmer lines. 

Conversely, the nebula around ULX-2 has a quasi-circular appearance and its line ratios and widths are consistent with a normal H {\sc ii} region, that is with stellar photo-ionization from one or two O-type dwarfs and a few blue supergiants. There is no radio counterpart and no other evidence of jet emission. The reason why some ULXs produce jets while others (with comparable X-ray luminosities) do not, both in the case of the two M\,51 sources and in the ULX class as a whole, remains unknown. The lack of high-ionization lines indicative of X-ray photo-ionization suggests that ULX-2 does not have a strong direct contribution to the ionization of the nebula. It is likely that the nebula is a chance alignment and is not powered by the ULX.

The X-ray and multiband flux from the two compact objects and surrounding nebulae is at the low end of the ULX class distribution; however, the peculiar property of both sources is that they show X-ray eclipses, from which one can in principle strongly constrain the binary parameters, the masses of the two compact objects, and the geometry of accretion and emission. Therefore, they are particularly suitable targets for follow-up multiband studies. 

The metallicity of the parent stellar population can have a direct effect on the type of ULXs and on their observational appearance. For example at super-solar metallicities, the ratio of neutron stars over black holes is strongly increased, which may be relevant to the ongoing debate over the relative proportion of neutron star and black hole accretors in the ULX population. Neutron stars with a Roche-lobe-filling massive donor star are also more likely to be eclipsing systems than black holes with the same type of donor star, which may explain why two eclipsing ULXs \citep{2016ApJ...831...56U} and another eclipsing X-ray binary with a luminosity just below the ULX threshold (Song et al.~, in prep.) were all found in this galaxy. Our LBT spectra of the ionized gas around ULX-2 are in agreement with the line ratios observed in previous optical studies of H {\sc II} regions in the disk of M\,51 (as discussed in Section 3.2.2); however, the conversion between such line ratios and a metallicity scale remains a subject of debate, with alternative models suggesting either $Z \approx 0.015$ (slightly sub-solar) or $Z \approx 0.04$ (2.5 times super-solar). Future, more accurate, determinations of the metallicity in nearby spiral galaxies will provide an important input parameter to test population synthesis models of ULXs.

\section{Acknowledgements}

We thank Gaelle Dumas for the reduced VLA+Effelsberg radio data. We also thank Jifeng Liu, Michela Mapelli, Mario Spera, Song Wang for useful discussions, and the anonymous referee for their constructive comments. RU acknowledges that this research is supported by an Australian Government Research Training Program (RTP) Scholarship. RS acknowledges support from a Curtin University Senior Research Fellowship; he is also grateful for support, discussions and hospitality at the Strasbourg Observatory during part of this work. JCAM-J is the recipient of an Australian Research Council Future Fellowship (FT140101082). The International Centre for Radio Astronomy Research is a joint venture between Curtin University and the University of Western Australia, funded by the state government of Western Australia and the joint venture partners. Based on observations made with the NASA/ESA Hubble Space Telescope, obtained from the Data Archive at the Space Telescope Science Institute, which is operated by the Association of Universities for Research in Astronomy, Inc., under NASA contract NAS 5-26555. These observations are associated with program 10452. This paper used data obtained with the MODS spectrographs built with funding from NSF grant AST-9987045 and the NSF Telescope System Instrumentation Program (TSIP), with additional funds from the Ohio Board of Regents and the Ohio State University Office of Research. IRAF is distributed by the National Optical Astronomy Observatories, which are operated by the Association of Universities for Research in Astronomy, Inc., under cooperative agreement with the National Science Foundation.
\label{lastpage}
\bibliography{references}
\end{document}